\documentclass{ar-1col}

\usepackage[colorlinks=false,pdfstartview=FitV,breaklinks=true]{hyperref}

\usepackage{bm}
\usepackage{graphicx}
\usepackage{multirow}
\usepackage{soul}
\usepackage{amsmath}
\usepackage{mathrsfs}
\usepackage{amssymb}
\usepackage{yfonts}
\usepackage{color}
\usepackage{xspace}
\usepackage{url}
\usepackage{verbatim}
\usepackage{mathtools}
\usepackage{upgreek}
\usepackage{units}
\usepackage{siunitx}
\usepackage{amstext}
\usepackage{booktabs}
\usepackage[sc]{mathpazo}
\usepackage{tabulary}
\usepackage{tabularx}
\usepackage{etoolbox}
\usepackage{ifoddpage}
\usepackage[version=4]{mhchem}
\usepackage[utf8]{inputenc}
\newcolumntype{Y}{>{\centering\arraybackslash}X}
\usepackage[official]{eurosym}

\newcommand{\qhat}{\hat{\mathbf{q}}}

\newcommand{\vlab}{\mathbf{v}_\mathrm{lab}}
\newcommand{\vmin}{v_{\rm min}}

\newcommand{\kms}{\textrm{ km s}^{-1}}

\newcommand{\dbd}[2]{\ifmmode \frac{\textrm{d}#1}{\textrm{d}#2}\else $\textrm{d}#1/\textrm{d}#2$\fi}
\newcommand{\pbp}[2]{\ifmmode \frac{\partial#1}{\partial#2}\else $\partial#1/\partial#2$\fi}

\newcommand{\ra}[1]{\renewcommand{\arraystretch}{#1}}

\newcommand{\vbf}{\mathbf{v}}

\DeclareMathAlphabet{\mathpzc}{OT1}{pzc}{m}{it}
\newcommand{\eV}{\text{e\kern-0.15ex V}\xspace}

\newcommand{\keVr}{\text{k\text{e\kern-0.15ex V}$_\mathrm{r}$}\xspace}
\newcommand{\keVee}{\text{k\text{e\kern-0.15ex V}$_\mathrm{ee}$}\xspace}

\newcommand{\TeV}{\text{T\kern-0.1ex \eV}\xspace}
\newcommand{\Gaia}{\!{\it Gaia}\xspace}
\newcommand{\cevns}{CE$\nu$NS\xspace}

\DeclareMathAlphabet{\mathpzc}{OT1}{pzc}{m}{it}

\newcommand{\Cygnus}{\textsc{Cygnus}\xspace}

\jname{Annual Review of Nuclear and Particle Science}
\jvol{71}
\jyear{2021}
\doi{10.1146/annurev-nucl-020821-035016}

\begin{document}

\markboth{Vahsen, O'Hare \& Loomba}{Directional recoil detection}

\title{Directional Recoil Detection}

\author{Sven E. Vahsen,$^1$ Ciaran A. J. O'Hare,$^2$ and\\ Dinesh Loomba$^3$
\affil{$^1$Department of Physics and Astronomy, University of Hawaii, Honolulu, Hawaii 96822, USA; email: sevahsen@hawaii.edu}
\affil{$^2$ARC Centre of Excellence for Dark Matter Particle Physics, The University of Sydney, School of Physics, NSW 2006, Australia; email: ciaran.ohare@sydney.edu.au}
\affil{$^3$Department of Physics and Astronomy, University of New Mexico, NM 87131, USA, email: dloomba@unm.edu}}

\begin{abstract}
Searches for dark matter-induced recoils have made impressive advances in the last few years. Yet the field is confronted by several outstanding problems. First, the inevitable background of solar neutrinos will soon inhibit the conclusive identification of many dark matter models. Second, and more fundamentally, current experiments have no practical way of confirming a detected signal’s galactic origin. The concept of directional detection addresses both of these issues while offering opportunities to study novel dark matter and neutrino-related physics. The concept remains experimentally challenging, but gas time projection chambers are an increasingly attractive option, and when properly configured, would allow directional measurements of both nuclear and electron recoils. In this review, we reassess the required detector performance and survey relevant technologies. Fortuitously, the highly-segmented detectors required to achieve good directionality also enable several fundamental and applied physics measurements. We comment on near-term challenges and how the field could be advanced.
\end{abstract}

\maketitle

\tableofcontents

\section{INTRODUCTION}

Over the last three decades, and despite substantial progress, direct evidence of interactions of galactic dark matter (DM) with Standard Model (SM) particles has been persistently lacking. To increase the probability of capturing a particle with very feeble interactions, detectors have typically advanced by lowering their energy thresholds, by reducing backgrounds, and by increasing their target masses up to the ton-scale, see References~\cite{Battaglieri:2017aum,Schumann:2019eaa} for reviews.
Now, the encroachment of the inevitable background of astrophysical neutrinos will prove to be the final---and in many cases, insurmountable---obstacle to the largest direct detection experiments ~\cite{Billard:2013qya}. This state of affairs has driven a resurgence in interest towards a comparatively little studied experimental technique.

It was first recognized by Spergel~\cite{Spergel:1987kx} that direct DM searches would be subject to a unique directional signature. The relative motion of the Solar System with respect to the Milky Way's DM halo should give rise to an anisotropic flux of DM with a peak incoming direction pointing back along the galactic plane, towards the constellation of Cygnus. 
A signal with a fixed galactic direction is not known to be mimicked by any cosmic or terrestrial background, and it is likely that any detected signal that was aligned in the direction opposing our galactic rotation would have to be related to the Milky Way's halo in some way. Moreover, unlike many other kinds of DM signals, which can vary considerably between experiments and particle candidates, the directionality of the flux is expected for almost all DM models~\cite{Kavanagh:2015jma, Catena:2015vpa}, and is highly robust against astrophysical uncertainties~\cite{OHare:2019qxc}. 

Directional detectors are uniquely capable of discriminating against the otherwise irreducible background of astrophysical neutrinos~\cite{Grothaus:2014hja, O'Hare:2015mda}---and a directional detector should generally enable the identification of a DM signal with far fewer events under any kind of background~\cite{Copi:1999pw, Morgan:2004ys, Billard:2009mf, Green:2010zm, Mayet:2016zxu, Vahsen:2020pzb}. 
Despite this strong motivation, directional detection is experimentally challenging, and the community, while growing, is still relative small. 

The majority of the experimental directional detection community has converged on the gas TPC as the optimum technology. The \Cygnus collaboration~\cite{Vahsen:2020pzb} has been formed from the convergence of several gas TPC collaborations who have run successful small-scale experiments in the past~\cite{Santos:2011kf,Baracchini:2020btb,Battat:2016xxe,Yakabe:2020rua,Vahsen:2011qx}. Gas TPC proponents have grown in number in recent years, and so has the readiness of many advanced readout technologies to detect keV-scale electron and nuclear recoils, discriminate between them, and reconstruct their directions. A large part of the inspiration for this progress has been the quest for dark matter, however a slew of other physics goals --- from measurements of neutrinos~\cite{Vahsen:2020pzb}, to fundamental and applied physics --- are also considered to be well-suited for a future large-scale gas TPC. 

Several excellent review articles on directional detection have been written in recent years~\cite{Ahlen:2009ev, Mayet:2016zxu,Battat:2016pap}. We highlight in particular Reference~\cite{Sciolla_2009}, which predates the other reviews, but provides additional valuable perspectives on select topics, including the Lindhard model for the energy loss of low-energy particles. 
More recently, this work has culminated in a ton-scale directional gas TPC design outlined as part of the \Cygnus project~\cite{Vahsen:2020pzb}.

Given that the motivation for a directional detector is growing, and that the experimental community is converging, it is timely to revisit the motivation and scope of directional recoil detection and carefully consider the present opportunities and remaining challenges. 
This review is structured from general to specific, and gradually transitions from an objective overview of the field into a more subjective presentation of the main challenges, ending with our personal view on optimal technologies for addressing these. In Section~\ref{sec:motivation} we introduce the diverse physics motivation for performing directionally sensitive recoil experiments. Then, in Section~\ref{sec:detectors} we describe the basic physics of recoils in gas targets, consider several broad technological approaches before listing specific examples of demonstrated or proposed detectors. We focus in on gas TPCs as the optimum approach for directional detection. In Section~\ref{sec:performance} we describe the required capabilities to achieve different physics goals, limiting ourselves mostly to gas TPCs. We find that TPCs with high-definition readouts (HD TPCs) can meet the performance requirements. In Section~\ref{hd_recoil_imaging} we illustrate physics measurements that can only be performed with such detectors. Finally, in Section~\ref{sec:summary}, we briefly summarize this review and present our recommendations for future work in the field.
\section{PHYSICS MOTIVATION}\label{sec:motivation}
\begin{figure}[t]
	\centering
	\includegraphics[width=0.955\textwidth]{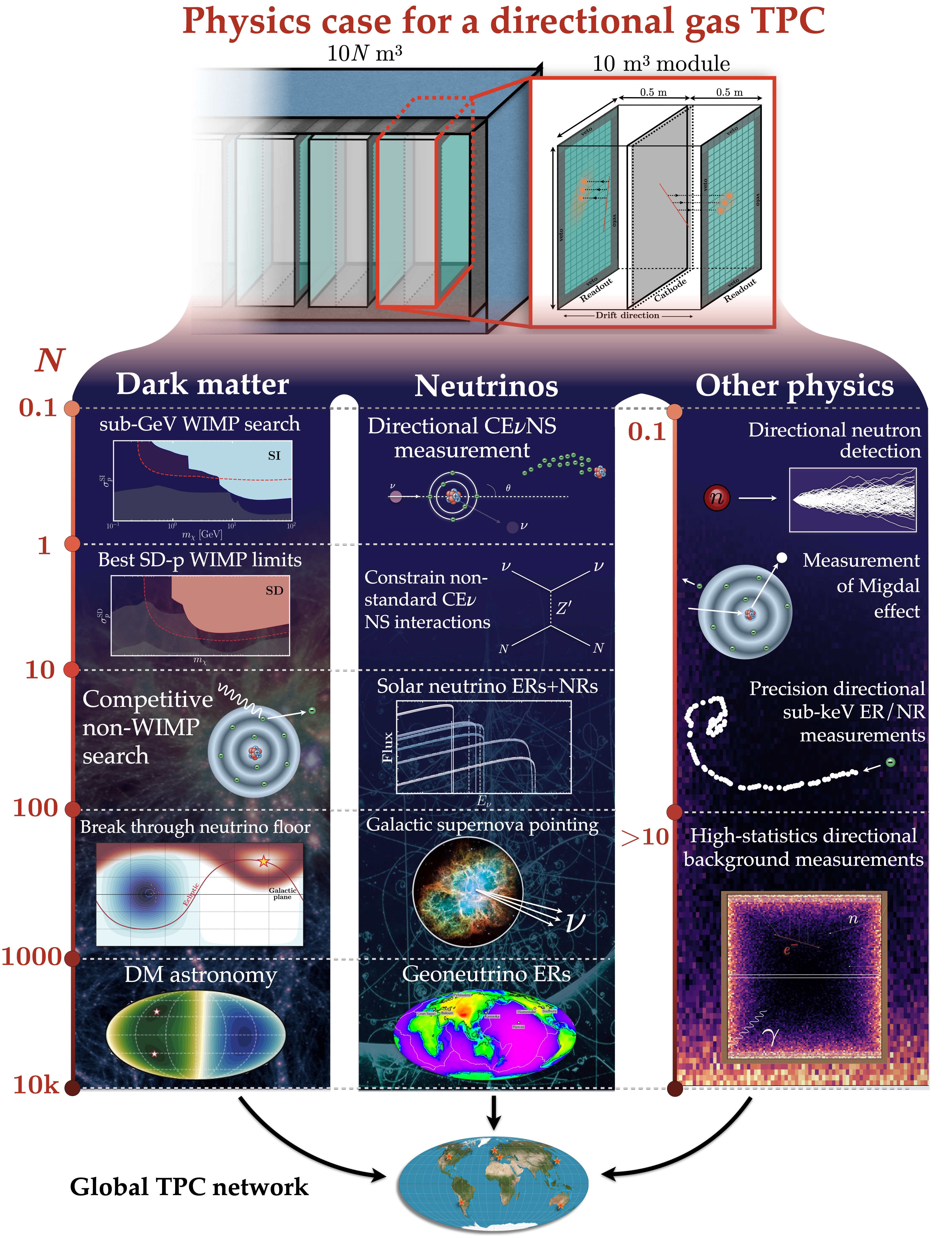}
	\caption{The physics case for a directional gas TPC, organized in terms of DM, neutrino physics, as well as other fundamental and applied physics. The cases are presented in order of the total size of a gas TPC experiment that would be needed, in terms of $N$, the number of 10 m$^3$ TPC modules operating close to atmospheric pressure. These volumes are not precise beyond an order of magnitude. We use a different scale for $N$ in the ``other physics'' column since these goals can be achieved with much smaller scale experiments.}
	\label{fig:summary}
\end{figure}

A summary of the physics case for a directional recoil detector is presented graphically in {\bf Figure~\ref{fig:summary}}. We will come back to this summary in Section~\ref{sec:summaryphysics} after we have discussed the full physics motivation for directional detection.

\subsection{Dark matter}
The search for DM remains the most compelling motivation for pursuing directional experiments. Let us recap why we believe DM signals to be generically directional. The commonly agreed-upon first approximation of a galaxy like our Milky Way is of a rotating disk embedded inside a spherical, isotropic, and non-rotating DM halo. Since we operate experiments in a reference frame that is moving at a velocity $\vlab$ with respect to the rest frame of the DM halo, the distribution of DM velocities that we observe, $f_{\rm lab}(\vbf)$ is obtained by boosting the galactic velocity distribution, $f_{\rm lab}(\mathbf{v},t) = f_{\rm gal}(\vbf + \vlab(t))$. 
Many of the characteristic signals of DM are due to this boost into our frame of reference. For instance, the time dependence of $\vlab(t)$ (due to the Earth-Sun relative motion) makes the flux modulate annually; and because the size of $|\vlab(t)|$ (due to the Sun-halo relative motion) is larger than the expected width of $f_{\rm gal}(|\vbf|)$, the flux will also be strongly anisotropic.

The anisotropy of the flux of DM particles is often touted as a smoking gun, resting on only a select few basic assumptions. This review is dedicated to assessing the feasibility of detecting such a signal experimentally. However, given that the entire field rests upon these assumptions, it is worth taking time to critically assess how confident we are in them.
\begin{summary}[Requirements for a directional DM signal that points back towards Cygnus]
\begin{enumerate}
    \item The local dark matter density, $\rho_0$ is nonzero.
    \item The solar velocity points along the galactic plane
    \item The DM halo is not co-rotating at a similar speed to galactic rotation
\end{enumerate}
\end{summary}
Firstly, the measurement of the density of unseen matter in the solar neighborhood has a long history that dates back to the work of Kapteyn in 1922~\cite{Kapteyn:1922zz}---predating even Zwicky's famous observations of the Coma cluster. The density of dark matter around us in the Milky Way can be inferred from the motions of stars, using them as tracers of the total gravitational potential. The inferred local DM density resulting from a variety of methods and datasets is typically $\rho_0\sim0.4$--$0.6$~GeV~cm$^{-3}$~\cite{deSalas:2020hbh}. These estimates are still heavily dominated by systematics but are, importantly, nonzero.

Secondly, the direction of the DM anisotropy towards Cygnus only relies on the assumption that the motion of the galactic disk points us in that direction. The solar velocity is of fundamental importance in galactic astronomy and astrometry in order to make sense of stellar parallaxes and proper motions, hence astronomers have conceived of numerous ways to precisely measure it~\cite{Bovy:2020}. The Solar System moves almost perfectly along the Galactic plane, at around $246 \pm 1 \, \kms$.
Even accounting for the aberration of the Earth's direction of motion due to its orbit around the Sun ($\sim 30\,\kms$) is not enough to cause the peak expected DM flux to ever point outside of the constellation of Cygnus.

The final assumption is also generally believed to be true, albeit with a slightly greater degree of uncertainty: the DM halo must not co-rotate with the galactic disk. If the DM halo did co-rotate, this would not doom all detection efforts, but it could substantially wash out directional signals. Triaxial halos like the Milky Way's~\cite{Iorio:2019} are formed hierarchically and will therefore typically have some angular momentum which would manifest as a figure rotation, or ``tumbling'', on Gyr timescales~\cite{Bryan:2007}. The figure rotation of the Milky Way has not been measured, but it could not be anomalously faster than the typical rotations seen in simulated Milky Way analogs, which are currently at the cusp of what could be observed via the influence on stellar streams~\cite{Valluri:2020lsq}. 
Put together, the assumption of a directional DM signal pointing back towards Cygnus seems rather robust.

\subsection{Directional signals of WIMP-like dark matter}\label{sec:wimps}
\begin{marginnote}[]
\entry{Weakly interacting massive particle (WIMP)}{a loosely defined particle candidate for DM that is usually assumed to be produced thermally in the early Universe.}
\end{marginnote}

The WIMP, supersymmetric or otherwise, remains a popular and widely-studied example of particle-like dark matter~\cite{Arcadi:2017kky}. The most common laboratory test of WIMP DM involves searching for their scattering with nuclei. The event rate of nuclear recoils as a function of recoil energy ($E_r$) and direction ($\qhat$) is given by integrating over the DM flux, $v f_{\rm lab}(\mathbf{v},t)$, multiplied by some differential scattering cross section $\textrm{d}\sigma/\textrm{d}E_r$ as follows, 
\begin{equation}\label{eq:WIMPRate}
\frac{\textrm{d}^2 R}{\textrm{d}E_r\textrm{d}\Omega_q}(E_r,t) = \frac{\rho_0}{2 \pi m_\chi m_N} \int_{v > v_\textrm{min}}{v^2 \delta(\vbf\cdot\qhat - \vmin) \, f_{\rm lab}(\mathbf{v},t) \frac{\textrm{d} \sigma}{\textrm{d}E_r}(E_r,v)} \, \textrm{d}^3 v \,.
\end{equation}
This formula will hold for all 2$\rightarrow$2 elastic nuclear scattering processes. We have also divided by the nuclear mass $m_N$ to get the event rate per unit detector mass. We only integrate over DM velocities kinematically permitted to produce a given recoil energy and direction. This consideration introduces both the low speed cutoff $v>v_{\rm min}(E_r)$, and the delta function, which enforces the non-relativistic kinematic relationship for the DM-nucleus scattering angle with respect to the initial DM velocity $\vbf$ (both defined in the lab frame),
\begin{equation}
 \frac{1}{v} \vbf \cdot \qhat = \cos{\theta} = \sqrt{\frac{m_N E_r}{2 v^2 \mu^2_{\chi N}}} = \frac{\vmin}{v} \, ,
\end{equation}
where $m_\chi$ is the DM mass, and $\mu_{\chi N}$ is the DM-nucleus reduced mass. 

The size of the DM cross section, and its dependence on recoil energy, DM velocity, particle identity, interaction type, and spin, are all model dependent. In general, this cross section is calculated from the squared matrix element, the transition probability for the DM-nucleus interaction. The most common treatment is to assume contact interactions which results in a cross section constructed from two operators, the identity ($\mathcal{O}_{\rm SI} = \mathbb{1}$) and one built from the DM and nuclear spins ($\mathcal{O}_{\rm SD} = \mathbf{S}_\chi \cdot \mathbf{S}_n$), often referred to as spin independent (SI) and spin dependent (SD), respectively. For SI and SD interactions, the matrix element introduces no additional $v$-dependence, meaning $\textrm{d}\sigma/\textrm{d}E_r \propto v^{-2}$. In these cases, the event rate inherits all of its direction dependence from the integral transform of the DM velocity distribution implied by Equation~(\ref{eq:WIMPRate}), which is known as the Radon transform~\cite{Gondolo:2002np}. 

\begin{figure}[t]
	\centering
	\includegraphics[width=0.97\textwidth]{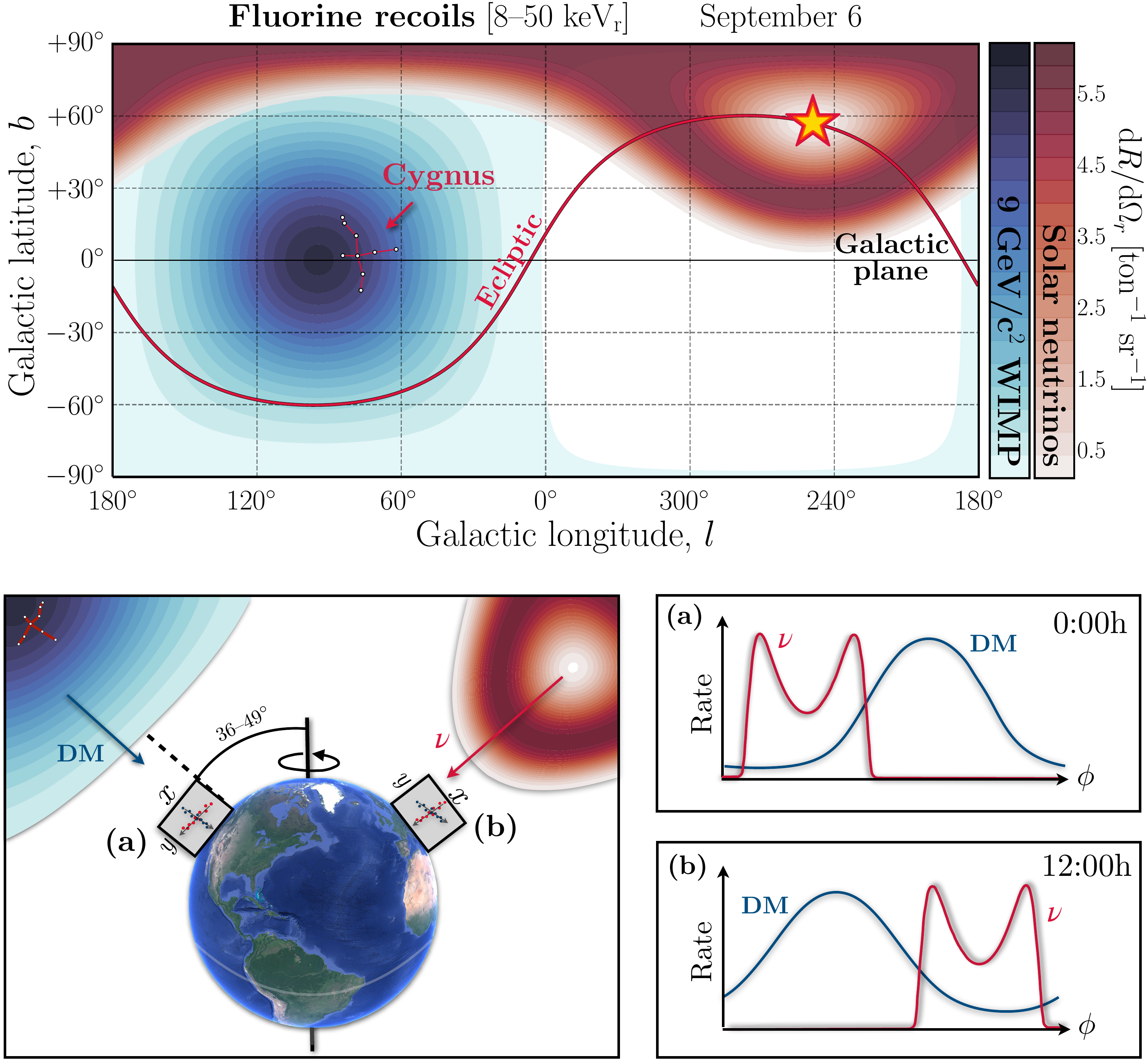}
	\caption{{\it Upper panel:}~the directional event rates from a 9 GeV DM particle (blue) and solar neutrinos (red) displayed in galactic coordinates $(l,b)$ in which the plane of the galaxy runs horizontally. We are moving towards the direction $(l,b) \approx (90^\circ,0^\circ)$, which means that this distribution of DM \emph{arrival} angles also peaks towards this direction. Solar neutrinos always originate from the ecliptic.  {\it Lower panels:}~in the coordinates of a detector at a fixed location on earth, the DM dipole translates into a directional oscillation. The Earth's rotation axis is tilted by an angle of 39--46$^\circ$ with respect to the galactic plane depending on date. We sketch typical event rates as a function of some angle $\phi$ on a 2d readout plane for two detectors separated by $180^\circ$ of longitude, or equivalently the same detector 12 hours later. Both the DM and the neutrino signals oscillate in angle over the day, but will always be separated from one another. Any local backgrounds will \emph{not} oscillate, so would be flat distributions in the lower right panels. Therefore this directional oscillation is also a powerful and experimentally observable signature of DM and neutrinos.}
	\label{fig:Skymaps}
\end{figure}
The angular nuclear recoil distribution from an $m_\chi = 9$~GeV WIMP undergoing SI scattering with $^{19}$F nuclei can be seen in the blue contours of {\bf Figure~\ref{fig:Skymaps}}. The event rate has been integrated over the range $E_r \in [8, 50]$~\keVr and remains roughly stationary in the galactic coordinates shown, given by longitude and latitude ($l,b$). The prominent dipole signature of DM-induced nuclear recoils is inherited from the boost of the velocity distribution, which is now centered on $-\vlab$. Since the Radon transform largely retains this directionality, the most probable recoil direction is also $\hat{\bf{q}} = -{\bf v}_{\rm lab}$. In contrast, opposing directions $\hat{\bf q} \approx + \vlab$, either have to come from the very high-speed tail of the velocity distribution, which is exponentially suppressed; or must have very large scattering angles, which have low recoil energies that are typically sub-threshold. This results in a very strong $\mathcal{O}(10)$ anisotropy in directions, if one takes the ratio between the integrated event rates in the two opposing hemispheres, and even higher if one selects smaller angles around $\qhat = \pm \hat{\vbf}_{\rm lab}$. The event rate will also become more strongly peaked towards Cygnus for higher recoil energies. This is because the low-speed cutoff for a given recoil energy, $\vmin$, increases with $E_r$. For higher energies, the only DM particles fast enough to scatter above $\vmin$ are those arriving from a head-on direction, aligning with $\vlab$.

The generic signal shown in {\bf Figure~\ref{fig:Skymaps}} is common to both $\mathcal{O}_{\rm SI}$ and $\mathcal{O}_{\rm SD}$ interaction operators. However these are not the only possible operators that could describe a DM-nucleus interaction. The most general effective field theory (EFT) construction of a non-relativistic DM-nucleus interaction could, in principle, incorporate any operators preserving Galilean, Hermitian, and time-reversal symmetries~\cite{Fan:2010gt, Fitzpatrick:2012ix, Anand:2013yka}. This results in a total of 15 terms (each, for protons and neutrons) built from combinations of momentum transfer, transverse velocity, DM spin, and nuclear spin operators.
In particular, operators that depend upon the DM transverse velocity, $v_{\perp}^{2}=v^{2}-q^{2}/4 \mu_{\chi N}^{2}$, introduce additional $v^2$-dependence not found in the SI and SD cross section expressions. These cases will introduce terms that depend upon the second moment of the Radon transform and lead to signals with additional ring-like features that slightly diminish the strength of the dipole~\cite{Kavanagh:2015jma,Catena:2015vpa}. Differentiating these kinds of features would be extremely difficult in nondirectional experiments.



\subsection{Directionality for dark matter discovery}
\vspace{1em}  \noindent {\bf Rejecting isotropic backgrounds.}
The strength of directional detection for DM discovery relies on the fact that no known backgrounds are believed to mimic (or even have any relation to) the directionality of a signal originating from the galactic halo. In fact, most backgrounds (with the notable exception of solar neutrinos) should be close to isotropic~\cite{Mei:2005gm}.
To get a feeling for the effectiveness of directional information for DM discovery, we can calculate a rough estimate for how many DM recoil directions would need to be measured to tell that the signal was \emph{not isotropic} (see References~\cite{Copi:1999pw, Morgan:2004ys, Copi:2002hm} for other approaches). To detect a dipole anisotropy, the most basic requirement would be a contrast in event numbers in the forward/backward hemispheres
 ($N_{\rm fw} - N_{\rm bw}$) greater than the typical 3$\sigma$ random deviation expected under isotropy, 3$\sqrt{N_{\rm fw}+N_{\rm bw}}$. Rearranging this requirement in terms of the event rates in each hemisphere ($R_{\rm fw, bw}$) gives the formula,
\begin{equation}
N_{\rm iso} \approx \left( 3 \, \frac{R_{\rm fw} + R_{\rm bw}}{R_{\rm fw} - R_{\rm bw}} \right)^2 \, .
\label{eq:reject_isotropy}\end{equation}
Taking the example of a $^{19}$F-based experiment with a $\sim$3~\keVr energy threshold and zero background, the number of events required to reject isotropy for DM masses $m_\chi = 10,\,100,\,1000$~GeV are $N_{\rm iso} \approx 12,\,16,\,17$. Fewer events are required for the lowest masses because all the recoils scattering above threshold are from the high-speed tail of the distribution, which is the most anisotropic part.
This simple non-parametric estimate already results in a promisingly low required number of events, however it is sensitive to isotropic background contamination. For example, if we assume signal events only make up a fraction $\lambda$ of the total number of events, this increases the value of $N_{\rm iso}$ by a factor $(1+1/\lambda)$, which could raise the required number for discovery up to $\mathcal{O}(100)$ for $\lambda\lesssim0.2$.

\vspace{1em}  \noindent {\bf Confirmation of a galactic signal.}
The test of isotropy presented above is highly simplistic since it reduces the signal down to only two angular bins. However, it reflects one of the key conceptual advantages of directional detection, which is that non-parametric statistical tests can be extremely powerful. More sophisticated tests described in the literature~\cite{Mayet:2016zxu} allow for unbinned recoil directions. These tests result in slightly smaller required event numbers, but importantly they do not require any additional modeling or assumptions beyond that of the background being roughly isotropic. Going one step further and confirming that the signal aligns with Cygnus requires around a factor of two more events~\cite{Green:2010zm}---still significantly smaller than the numbers of events required to make a similar statement with a non-directional experiment. 
Nonetheless, modeling the signal and background would still be the most desirable strategy in practice. This would allow the kinematic correlation between recoil direction and energy to be included, and would result in even lower required numbers to point towards Cygnus, at the cost of more model dependence~\cite{Billard:2009mf}. 

\vspace{1em}  \noindent {\bf DM discovery via sidereal modulation.}
So far we have assumed that the galactic dipole is a measureable signal. This implicitly assumes that all three components of each recoil vector, including its sign or ``head/tail'' can be measured. Actual detectors, however, may not be sensitive to all three components, and obtaining a head/tail signature is often particularly challenging. A lack of complete three-dimensional recoil vectors implies that individual events cannot be unambiguously rotated into galactic coordinates, and any alignment with Cygnus cannot be checked. This would seem problematic when thinking about the simplistic DM discovery arguments we have presented so far. 
\begin{marginnote}[]
\entry{Head/tail sensitivity}{a nuclear recoil detector's ability to distinguish between a recoil direction vector $\qhat$ and the opposite vector $-\qhat$.}\end{marginnote}
\begin{marginnote}[]
\entry{Sidereal day}{a measure of the Earth's rotational period with respect to the fixed stars. In contrast to the solar day which is measured with respect to the Sun.}
\end{marginnote}

Even in these cases, however, a form of directional discovery is possible. This alternative strategy relies on the rotation of the Earth to fill the gap in information, as depicted in the lower panels of {\bf Figure~\ref{fig:Skymaps}}. For example, if only a 2d projection of each recoil direction were measurable, then the projected dipole signature would rotate over the course of one sidereal day. A signal that modulated with the sidereal day, by definition, would have to be of galactic origin and unrelated to the Earth-Sun system. Any local systematic effects exhibiting a daily modulation (for instance with temperature) would presumably have to follow the solar day. Accounting for this diurnal variation and contrasting it with the background can allow 2d and 1d experiments to regain sensitivity~\cite{Billard:2014ewa}. 

\begin{figure}[t]
  \checkoddpage
  \edef\side{\ifoddpage l\else r\fi}%
  \makebox[\textwidth][\side]{%
    \begin{minipage}[t]{1.2\textwidth}
      \centering
      \includegraphics[width=0.97\textwidth]{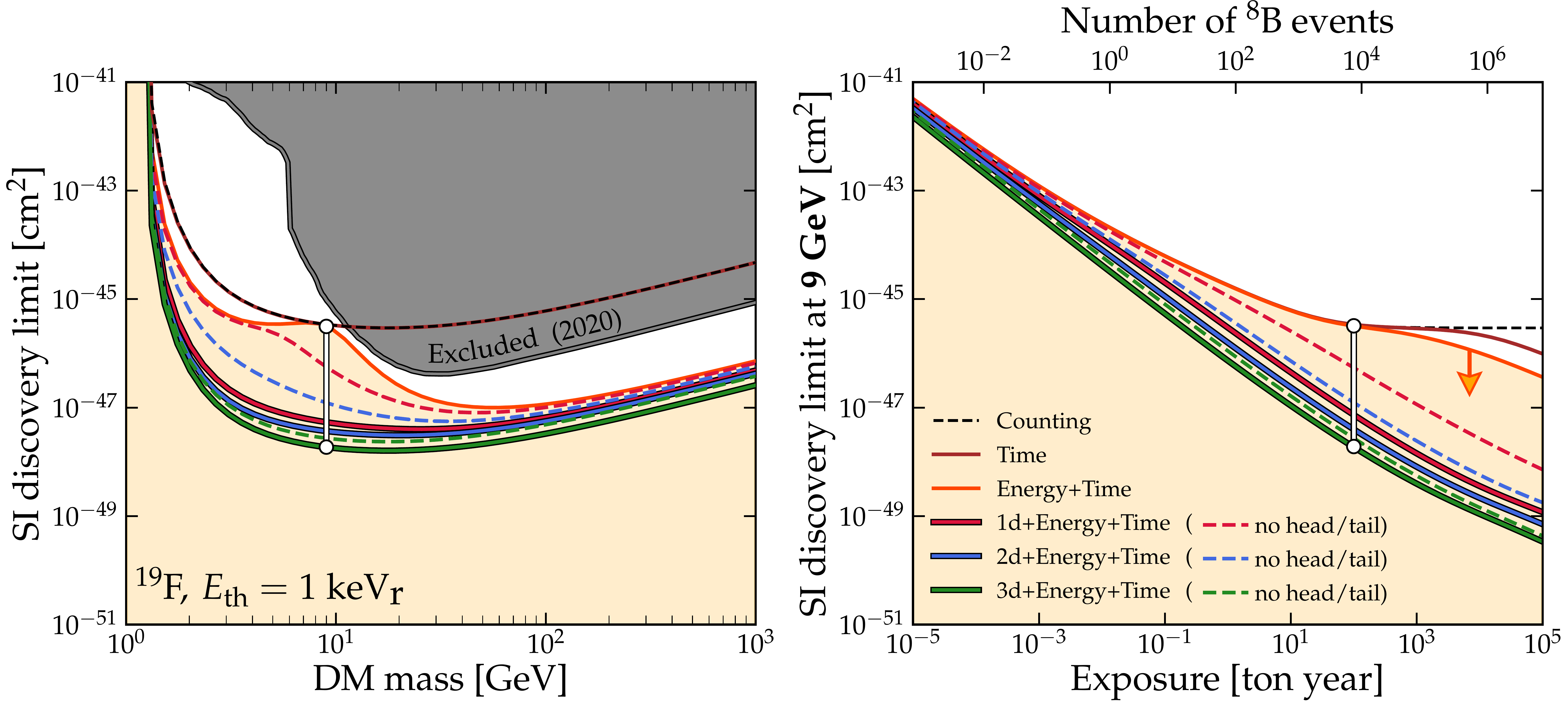}
          \caption{Effect of directionality and other detector capabilities when setting limits on DM cross sections in the presence of neutrino background. Left: discovery limits versus DM mass for a fixed detector exposure. Right: discovery limits versus  exposure for a fixed DM mass. The vertical white lines show where the two panels intersect. In both panels, the only difference between the different lines is the information that is used in the analysis. The lowest line (green) uses all information available (green: 3d directionality, recoil energy, and event time) whereas the highest line assumes the most minimal amount of information possible (black: the number of events only). We shade underneath the orange curve to highlight the range of WIMP models that are inaccessible without directional information for a given exposure and target.}\label{fig:NuFloor}
    \end{minipage}%
  }%
\end{figure}

\vspace{1em}  \noindent {\bf DM discovery under the neutrino background.}
Ultimately, the best prospects for discovery and characterization of a signal will be achieved when all recoil direction, time, and energy information are incorporated into a complete model. This is most clearly demonstrated when the dominant background is not isotropic; and there turns out to be a highly notable example: the solar neutrino background. The keV-scale nuclear recoils from coherent neutrino-nucleus elastic scattering (\cevns) of solar neutrinos will be the most problematic background for the upcoming generation of WIMP searches. The \cevns background from 
$^{8}$B neutrinos, is particularly troublesome because the resulting spectrum of nuclear recoil energies resembles that of DM for $m_\chi = $~5--10 GeV. In fact, spectral matching occurs at many other DM masses that each overlap with different fluxes of neutrino. This mimicry of the DM signal by an otherwise irreducible background is what gives rise to the well-known ``neutrino floor''~\cite{Billard:2013qya}. Without substantial improvements to the already precisely known neutrino fluxes, progress of direct DM detection towards smaller cross sections will be limited for the next decade and beyond.

Directionality is an attractive prospect for circumventing the neutrino floor because the distinct angular distributions of DM and solar neutrinos recoils allows for optimum discrimination between the two~\cite{Grothaus:2014hja, O'Hare:2015mda}, see {\bf Figure~\ref{fig:Skymaps}}. The angular recoil distributions are distinct due to the separation between the path of the Sun (the ecliptic) and the constellation Cygnus across the sky. 
We display a quantitative demonstration of this in {\bf Figure~\ref{fig:NuFloor}}. In both panels, we show the discovery limits (defined as the median cross section that could be detected at 3$\sigma$) for $^{19}$F-based experiments with a 1~\keVr energy threshold and ton-scale target masses.
The left-hand panel fixes the exposure (100 ton-year) but shows the limit as a function of DM mass, whereas the right-hand panel fixes the DM mass (9 GeV) but shows the limit versus exposure. The most important result shown by this figure is that the directional limits (green, red, blue) scale almost as $\sim 1/\textrm{Exposure}$, whereas the nondirectional limits all scale as (at best) $1/\sqrt{\textrm{Exposure}}$. 


\subsection{Directional signals for probing beyond-WIMP dark matter}
Directionality is a broadly model-independent prediction that should be present in most DM models. 
We highlight examples of non-WIMP models where directional detectors appear better suited than their non-directional counterparts for detection or model characterization.

\vspace{1em}  \noindent {\bf Modified DM-nucleus kinematics.}
One of longest-standing elaborations on the WIMP was proposed almost two decades ago~\cite{TuckerSmith:2001hy}, but is still the subject of investigation~\cite{Eby:2019mgs,Zurowski:2020dxe,Bramante:2016rdh}. So-called ``inelastic DM'' models introduce an excited state for the DM particle that it can either be excited to, or deexcited from, during a collision with a nucleus. These models modify the formula for $\vmin$ to account for the mass splitting between these states. In inelastic DM models with an available excited state, the distribution of recoils would be more focused towards Cygnus since slower particles would not be able to scatter with enough energy to get excited. Such a signal would be much more readily observed in the angular distribution~\cite{Finkbeiner:2009ug}. So untangling the mass spectrum of DM and distinguishing elastic from inelastic interactions would be a unique advantage of a directional detector~\cite{Lisanti:2009vy}. 

\vspace{1em}  \noindent {\bf DM in detectors with large volumes.}
For some DM models, certain kinds of directional recoil detectors are attractive, but not for their directional sensitivity. Directional recoil measurements typically prefer lower-density targets, hence  directional detectors will typically require large total volumes to reach competitive exposures. It turns out that the DM event rate for certain classes of model scales with the geometric \emph{size} of the detector rather than the total mass. One example of this is when the DM is strongly interacting and extremely heavy. In these models, the flux of particles is low, but if one does cross the detector, the probability of it generating multiple scattering events is very high~\cite{Bramante:2018qbc,Bramante:2019yss}. In this case, the number of events would scale with the cross sectional area of the experiment. This was studied recently in Reference~\cite{Clark:2020mna} considering the reach of the 1 m$^2$-scale of Xenon1T to DM masses up to $10^{18}$~GeV. Since the masses are so high, the momentum imparted in each scattering event is negligible compared to the DM's kinetic energy. This means the multiple scatters would be essentially colinear and would be even more anisotropically distributed than WIMPs. 
Another instance of this is the case of ``luminous DM''~\cite{Feldstein:2010su} which is related to inelastic DM but has the added feature of electromagnetic emission from the decay of the excited DM state. This idea was studied recently in the context of large-scale directionally sensitive detectors~\cite{Eby:2019mgs}. To gain novel sensitivity, the experiment would need to be equipped with photodetectors to identify the scintillation emitted when a DM particle that was excited in a prior interaction inside the Earth then decays inside the detector volume. 

\vspace{1em}  \noindent {\bf Fluxes of DM from other directions.}
The flow of DM from Cygnus is a robust prediction, however in some models this population of particles may simply be impossible to detect, especially if they are much lighter than the typical GeV-scale WIMP. However, many have wondered if a sub-population of these light particles could be boosted to detectable energies. Such scenarios need not be contrived into existence, but could be a guaranteed prediction of certain models and the primary way they would be detectable. One such scenario is ``cosmic ray upscattered DM''~\cite{Bringmann:2018cvk, Alvey:2019zaa, Dent:2019krz, Guo:2020oum, Dent:2020syp}. The DM in this case would be a standard WIMP-like particle, however the upscattering by GeV cosmic rays in the galaxy allows experiments to reclaim sensitivity to sub-GeV masses that would normally generate signals well below threshold. The flux of upscattered DM would inherit directionality from the spatial distribution of DM in the galaxy as well as some of the directionality of cosmic rays in the interstellar medium~\cite{Guo:2020oum}. The large cross sections of models for which this effect is relevant also means that the DM will be attenuated noticeably by the Earth~\cite{Bringmann:2018cvk}. This will cause a daily modulation of the flux and will suppress upward-going DM arrival directions. 

Another boosted population of DM was suggested recently, specifically in the context of directional recoil detectors~\cite{Baracchini:2020owr}. The model in question involves an MeV-scale particle which could represent a viable light DM candidate~\cite{DeRocco:2019jti} but would also be generated in abundance during supernovae. The diffuse flux of semi-relativistic particles from galactic supernovae would generate nuclear recoil signals comparable in energy to a cold population of GeV-scale WIMPs. However, this flux should peak towards the Galactic center, around $90^\circ$ away from the expected DM flux. The discrimination of these two fluxes is only possible with a directional experiment. 

\vspace{1em}  \noindent {\bf Electron recoils.}
Finally, directional detectors present novel opportunities for probing non-WIMP models via electron recoils. Bosonic DM candidates, such as dark photons~\cite{An:2014twa} and axion-like particles~\cite{Jaeckel:2010ni}, could undergo absorption processes in atoms, resulting in the emission of electrons with energies equal to the DM mass~\cite{Derevianko:2010kz}. Therefore electrons from keV-scale mass particles are readily observable in most DM searches. The key issue is how to separate these signal electrons from all other sources of electron recoils. A major advantage of directional detectors in this context is the ability to not just discriminate electrons from nuclear recoils, but to discriminate many sources of electron recoil from each other. 

\subsection{Directional signals of the dark matter halo}
An isotropic Gaussian is a convenient, first approximation for the DM velocity distribution $f(\vbf)$. Several pieces of observational evidence, however, already suggest it may be inaccurate in specific ways~\cite{Evans:2018bqy}. Uncertainty in the velocity distribution reduces the reliability of predicted direct detection event rates. Understanding these astrophysical uncertainties is important when setting limits on DM, but becomes essential when measuring DM properties. This is where directional measurements become extremely useful. Incorporating directional information allows for far superior measurements of DM particle properties while also mitigating astrophysical uncertainties~\cite{Lee:2012pf, OHare:2014nxd,Kavanagh:2016xfi}.

The velocity distribution in itself is of great post-discovery interest. While the Gaussian distribution of the SHM is monolithic, the real velocity distribution will likely possess complexity and substructure acquired over the Milky Way's tumultuous 13 Gyr lifetime~\cite{Necib:2018iwb, Evans:2018bqy,Bozorgnia:2019mjk,Necib:2019zka}. Such substructure has already been revealed in the dark halo's stellar counterpart~\cite{Kr18, Be18, HelmiReview, Naidu2020} by the revolutionary dataset from the \Gaia mission~\cite{GaiaDR2}. Some of the most relevant substructures will be in the form of tidal streams and unmixed debris, which are generic predictions of hierarchical structure formation and have been observed abundantly in the Milky Way's outer halo already. DM streams are possibly the most exciting form of substructure for directional detectors since they are kinematically localized around a single incoming direction~\cite{OHare:2014nxd}. The signature of this kind of substructure would be almost invisible in the recoil energy spectrum, but very prominent in the angular spectrum~\cite{OHare:2018trr,Adhikari:2020gxw}.  Certainly, any direct measurement of this structure in DM experiments would be of profound importance to galactic astronomy.

\subsection{Directional signals of neutrinos}\label{sec:neutrinos}
Experiments seeking direct signals of WIMP-like DM can naturally serve a dual-purpose as detectors of astrophysical or terrestrial sources of neutrinos and serve a diverse catalog of potentially novel physics. Since DM recoil detectors are optimized to detect $\mathcal{O}(1$--$100)$ keV recoil energies and require at least $\mathcal{O}(\textrm{few})$ events per ton-year of detector mass, the natural sources that are realistically within reach are (in order of detectability) Solar~\cite{Billard:2013qya}, nearby galactic supernovae~\cite{Lang:2016zhv} and geological~\cite{Leyton_geo} neutrinos. Artificial fluxes of neutrinos could also be measured if a DM detector is placed near a neutrino source, such as a beamline, beam dump, or a nuclear reactor.

Recoil detectors are sensitive to both coherent neutrino-nucleus elastic scattering (\cevns\footnote{pronounced ``sevens''}) and neutrino-electron elastic scattering. While the latter is a valuable channel for probing astrophysical neutrinos~\cite{Tomas:2003xn}, \cevns has so far only been measured by COHERENT using a stopped pion neutrino source~\cite{Akimov:2017ade, Akimov:2020pdx}. The elastic scattering angle between the neutrino direction and the recoil direction for a particle of mass $m$, is~\cite{Vogel:1989iv},
\begin{equation}\label{eq:nuscatteringangle}
 \cos{\theta} = \frac{E_\nu + m}{E_\nu}\sqrt{\frac{E_r}{E_r+2 m}} \, .
\end{equation}
The neutrino-electron and neutrino-nucleus recoils will generally be well-correlated with the original neutrino direction. 
\cevns is a flavor-blind interaction proceeding via a neutral current, and at low momentum transfer is coherently enhanced by a factor that depends approximately on the number of neutrons in the target nucleus~\cite{Freedman:1973yd}. Neutrino-electron scattering, on the other hand, has both charged and neutral current contributions and the cross sections for $\nu_e$ and $\bar{\nu}_e$ are the highest by almost an order of magnitude.

\begin{table}[t]\centering
\ra{1.3}
\caption{Approximate expected numbers of neutrino-induced nuclear and electron recoils assuming a 1000 m$^3$ target volume, 1 atmosphere pressure, and an exposure time of 1 year.}\label{tab:nurates}
\begin{tabularx}{\textwidth}{l|YYY|YYY|YYY}
\hline\hline
{\bf Nuclear recoils} & \multicolumn{3}{c|}{SF$_6$} & \multicolumn{3}{c|}{CF$_4$} & \multicolumn{3}{c}{He}\\
Threshold [\keVr] & 1  & 5  & 10  & 1  & 5  & 10  & 1  & 5  & 10 \\
\hline
Solar (mainly $^8$B) & 73  & 15  & 2  & 54  & 16  & 3  & 3  & 2  & 1 \\
3 kpc supernova  & 25  & 18  & 12  & 18  & 13  & 10  & 0.6  & 0.5  & 0.5 \\
\hline \hline
 \multicolumn{10}{c}{}\\
  \hline \hline
{\bf Electron recoils} & \multicolumn{3}{c|}{SF$_6$} & \multicolumn{3}{c|}{CF$_4$} & \multicolumn{3}{c}{He}\\
Threshold [keV] & 5  & 500  & 1000  & 5  & 500  & 1000  & 1  & 500  & 1000 \\
\hline
Solar (Total) & 537  & 42  & 4  & 438  & 34  & 3  & 102  & 8  & 0.8 \\
Solar (CNO)  & 15  & 5  & 0.6  & 12  & 4  & 0.5  & 3  & 0.9 & 0.1\\
Geoneutrinos  & 0.2  & $<$0.1  & $<$0.1   & 0.2  & $<$0.1  & $<$0.1   & $<$0.1  & $<$0.1   & $<$0.1  \\
\hline \hline
\end{tabularx}
\end{table}

\vspace{1em}  \noindent {\bf Solar neutrinos.}
The event rates for the most relevant sources of neutrino-induced nuclear and electron recoils are shown in Table~\ref{tab:nurates} for a range of thresholds and possible target gases. For both nuclear and electron recoils, the dominant natural source of neutrinos for a DM recoil experiment will be the Sun. The Sun produces several well-understood fluxes of neutrinos from a variety of processes involved in nuclear fusion. Most \cevns recoils will be from the $E_\nu\sim10$~MeV neutrinos from the decay of $^8$B nuclei. These are not the highest energy neutrinos emitted by the Sun---those being the neutrinos from $^3$He-proton fusion---but they are the only ones that can generate a sizeable rate of nuclear recoils at keV energies. 
For electron recoils however, the kinematics result in much higher recoil energies at constant neutrino energy than in the case of nuclear recoils. This makes the electron recoil signature a very promising target for the directional detection community. In this case, the most substantial contribution will be from $pp$ fusion which generates the vast majority of the total solar flux.

Unfortunately, $pp$ and $^8$B neutrinos are not the most interesting type of solar neutrino astrophysically, since both fluxes are known rather precisely~\cite{Vinyoles:2016djt, Bergstrom:2016cbh}. Instead, one of the most sought-after solar fluxes are the neutrinos emitted in the Sun's ``CNO cycle''. Three fluxes of neutrinos labeled, $^{13}$N, $^{15}$O and $^{17}$F, have only just been observed by Borexino after a heroic background modeling effort~\cite{Agostini:2020mfq}. CNO neutrinos are almost entirely hidden under backgrounds, both from their fellow and more abundant solar neutrinos, as well as from radioactive contaminants. Yet they are a highly prized signal from a solar physics standpoint. A firm measurement of the CNO flux would help understand a long-standing disagreement between two models for the Sun's heavy element content~\cite{Villante:2019tcd}. This quantitative issue is subtle but has far-reaching consequences for astronomy since almost all determinations of astronomical elemental abundances rely upon the solar abundances.

The measurement of low energy solar neutrinos via directional electron recoils is, surprisingly, not a new idea. Largely-forgotten work from the 1990s~\cite{Seguinot:1992zu,Arpesella:1996uc}, proposed the use of a TPC filled with high densities of gases like He and CF$_4$ to detect solar-neutrino electron recoils $\gtrsim$100 keV. While most fluxes generating high numbers of electron recoils are now well-measured, the detection of CNO neutrinos is an intriguing possibility. 
The most obvious novel aspect of directionality is background rejection. Unfortunately, in the case of CNO neutrinos, the major backgrounds will be \emph{other} solar neutrinos. However, directionality is novel in another way when dealing with a signal originating from a single direction. Given the known position of the Sun and the combined measurement of recoil energy and direction, in theory, this information permits event-by-event reconstruction of the neutrino energy spectrum. 
A modern gas TPC with a 1000 m$^3$ volume at atmospheric pressure or higher could make  \emph{directional} measurements down to $\mathcal{O}(10)$~keV energies, much lower than the current threshold of Borexino of $\sim$160 keV. 

\vspace{1em}  \noindent {\bf Geoneutrinos.}
Antineutrinos from the Earth have very low energies, $\lesssim 4.5$~MeV, hence electron recoils are required for detection. Fluxes are very low, meaning 100 to 1000 ton-year exposures are needed to make a scientifically useful observation~\cite{Leyton_geo}. Directional detectors would enable rejection of solar neutrino backgrounds, and utilizing elastic scattering would enable sentivitity to lower neutrino energies than experiments like KamLAND~\cite{Araki:2005qa} and Borexino~\cite{Bellini:2010hy}, which rely on capture via inverse beta decay. Crucial components of lower energy geoneutrino sources like $^{40}$K nuclei have gone undetected because of the 1.8 MeV threshold of inverse beta decay. A measurement of these sources could help constrain the radioactive contribution to the Earth's surface heat flow~\cite{se-1-5-2010, Gando:1900zz}. A 10 ton-scale detector operating for 10 years would be capable of a 95\%~CL measurement of the $^{40}$K flux~\cite{Leyton_geo} and go some way to understanding this problem. 

\vspace{1em}  \noindent {\bf Galactic supernovae.}
The $\sim$10~MeV energies of neutrinos from supernovae make them a prime target for DM detectors~\cite{Lang:2016zhv}. A detection via \cevns would probe the all-flavor burst flux, thereby providing a normalizing measure of the total luminosity, which could be compared against flavor-dependent measurements made by other neutrino observatories. For a 1000 m$^3$ gas experiment, the nuclear recoil event rate within a $\sim$10~s burst window could be similar to a year's worth of solar neutrinos as long as the supernova occurred within around 3~kpc of the Earth (the average galactic supernova distance is estimated to be around 10 kpc~\cite{Mirizzi:2006xx}). The main advantage of directionality for the detection of supernova neutrinos via \cevns is pointing, which would potentially provide a valuable service to follow-up electromagnetic observations~\cite{Kharusi:2020ovw}.

\vspace{1em}  \noindent {\bf Artificial neutrinos.}
Physics possibilities with an artificial-neutrino-source \cevns experiment are extensive. If an experiment were placed close enough to a nuclear reactor, it would enjoy a generous flux of $\bar{\nu}_e$. Stopped pion sources are also available and were recently used by COHERENT for the first measurement of \cevns~\cite{Akimov:2017ade, Akimov:2020pdx}. A potentially more fruitful application of directional detectors would be to operate near a beam dump. In such a setup, even a small-scale gas TPC could make the first directional measurement of \cevns. This idea is being pursued for $\nu$BDX-DRIFT~\cite{nuBDX-DRIFT}, a proposal to place a negative-ion TPC behind the NuMI proton beam dump at Fermilab, with the longer-term goal of operating a TPC at the DUNE Near Detector Complex. Early estimates suggest that a 1 m$^3$ TPC could achieve a low-background directional measurement of \cevns with around a year of operation. 

\vspace{1em}  \noindent {\bf Beyond-the-SM neutrino interactions.}
Measurements using artificial neutrino sources such as reactor, stopped pions, or beam dumps all offer a potential gateway to beyond-the-SM physics measurements. These could include the detection of up-scattered heavy neutrinos, axion-like particles~\cite{Brdar:2020dpr, Dent:2019ueq}, and light DM candidates~\cite{Dutta:2019nbn}, which may produce novel signatures in angular spectra. With even higher statistics, constraining and disentangling a wide range of additional mediators that could be involved in \cevns could also greatly benefit from additional information present in the angular distribution~\cite{Abdullah:2020iiv}. Though the measured \cevns cross section is consistent with the SM, there is still room for beyond-the-SM corrections below experimental bounds~\cite{AristizabalSierra:2019ykk}. In the context of DM detectors, the effects of new mediators taking part in \cevns have been considered, for example, in References~\cite{Cerdeno:2016sfi, Bertuzzo:2017tuf,Boehm:2018sux, AristizabalSierra:2017joc, Boehm:2020ltd}. As well as providing opportunities for discovery, the added uncertainty in the \cevns background also presents problems for conventional recoil detectors. As we discussed earlier, the height of the neutrino floor is controlled by the neutrino event rate, and its uncertainty. Non-standard interactions and additional mediators have the potential to increase both. In particular, the event rate at low energies relevant for GeV and sub-GeV WIMP searches is precisely where there is substantial room for large deviations from the SM. Conducting a directional search to unravel these subtleties and distinguish them from a potential DM signal, is therefore even more warranted.

\subsection{Summary of the physics case for a directional recoil detector}\label{sec:summaryphysics}
{\bf Figure~\ref{fig:summary}} summarizes the rich physics potential of a directional recoil experiment, assuming it is realized as a gas TPC. 

\vspace{1em}  \noindent {\bf Dark matter.} Directional DM searches with gas TPCs have reached the 1~m$^3$ scale, but most have higher energy thresholds than desirable. Achieving improved directional sensitivity (Section~\ref{sec:angularperformance}) and particle identification (Section~\ref{sec:particleID}) should be feasible in the near future. Hence, a competitive low mass DM search with ultralow threshold is a natural first goal for a program of gas TPCs. The next goal---setting competitive SD WIMP limits---is also a natural one for TPCs using fluorine-based gases. A 10~m$^3$ detector should be sufficient to produce the world's best SD cross-section limits~\cite{Vahsen:2020pzb}. 

\vspace{1em}  \noindent {\bf Neutrinos.} The first directional measurement of \cevns should be possible in the near future with a small-scale TPC placed near to a neutrino source. By 1000~m$^3$, an atmospheric pressure TPC should already see in excess of $\mathcal{O}(10)$ nuclear recoils and $\mathcal{O}(100)$ electrons from solar neutrinos every year. With larger volumes, it may be possible to point to galactic supernovae out to 10~kpc, or even study the angular distribution of geoneutrinos.

\vspace{1em}  \noindent {\bf Other physics.} To reach the large volumes required to perform DM and neutrino physics suggested above, it will first be necessary to demonstrate the performance at smaller scales.
There is a host of demonstrated and proposed applications of smaller-scale TPCs that will enable a rich research program. Small gas TPCs are already operating as directional neutron background detectors~\cite{Jaegle:2019jpx}. Other proposed applications include topics as diverse as neutron imaging, passive detection of special nuclear material, fuel rod monitoring, and medical physics. TPCs with HD readout, specifically, are also uniquely promising for verifying the physics of low-energy nuclear recoils, which all DM experiments rely on, but which is not well constrained. For example, it may be possible to perform a direct verification of the Migdal effect. Due to the increasing relevance of this effect to searches for DM, this may be one of the most interesting immediate physics goals for smaller-scale TPCs. 

\section{DETECTING RECOIL DIRECTIONS}\label{sec:detectors}
\begin{figure}[t]
  \checkoddpage
  \edef\side{\ifoddpage l\else r\fi}%
  \makebox[\textwidth][\side]{%
    \begin{minipage}[t]{1.2\textwidth}
\begin{minipage}{.33\textwidth}
  \centering
  \textbf{Fluorine recoil}
  \includegraphics[trim = 5mm 0mm 12mm 0mm, clip,width=0.99\textwidth]{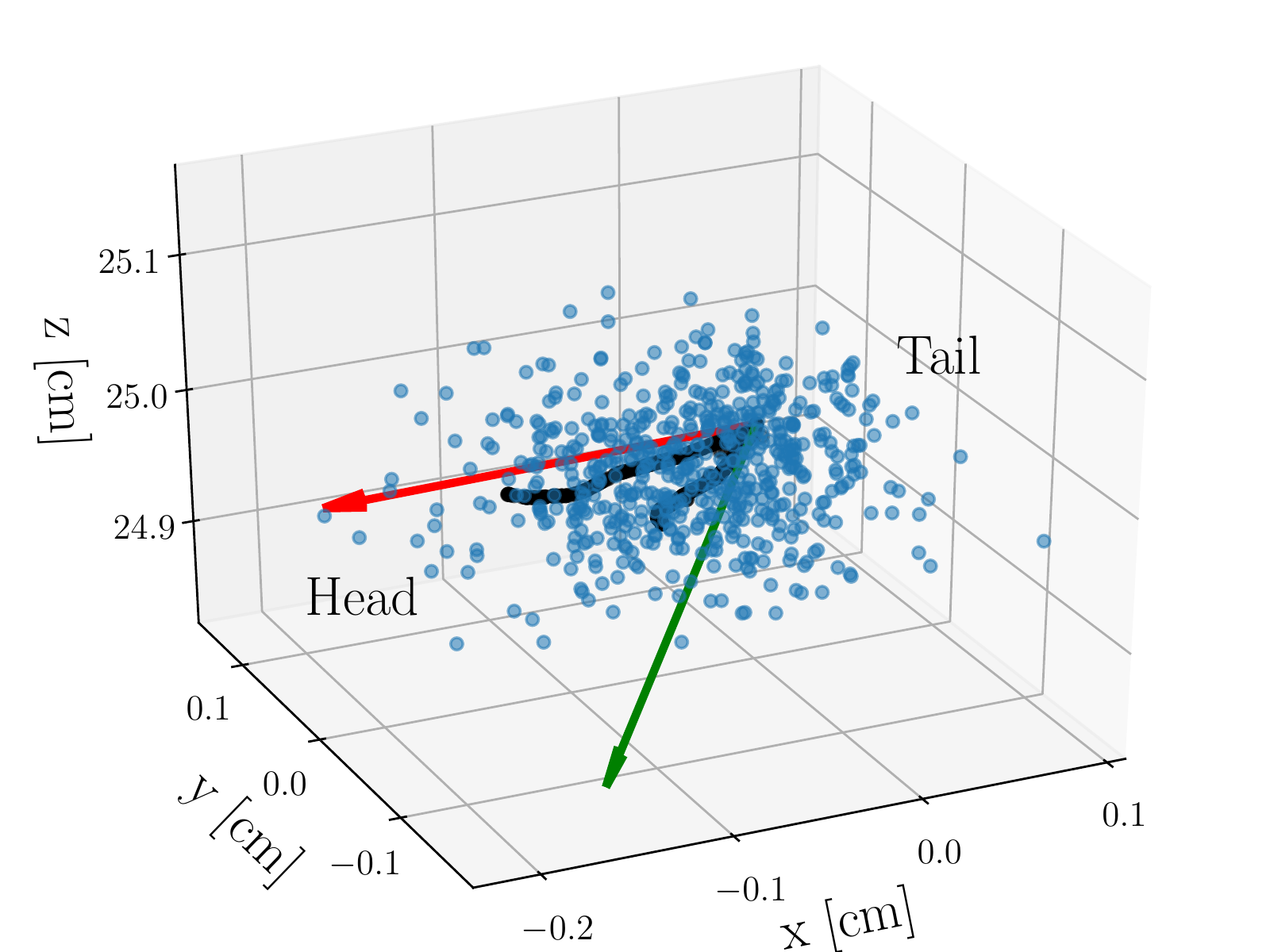}
\end{minipage}%
\begin{minipage}{.33\textwidth}
  \centering
  \textbf{Helium recoil}
  \includegraphics[trim = 5mm 0mm 12mm 0mm, clip,width=0.99\textwidth]{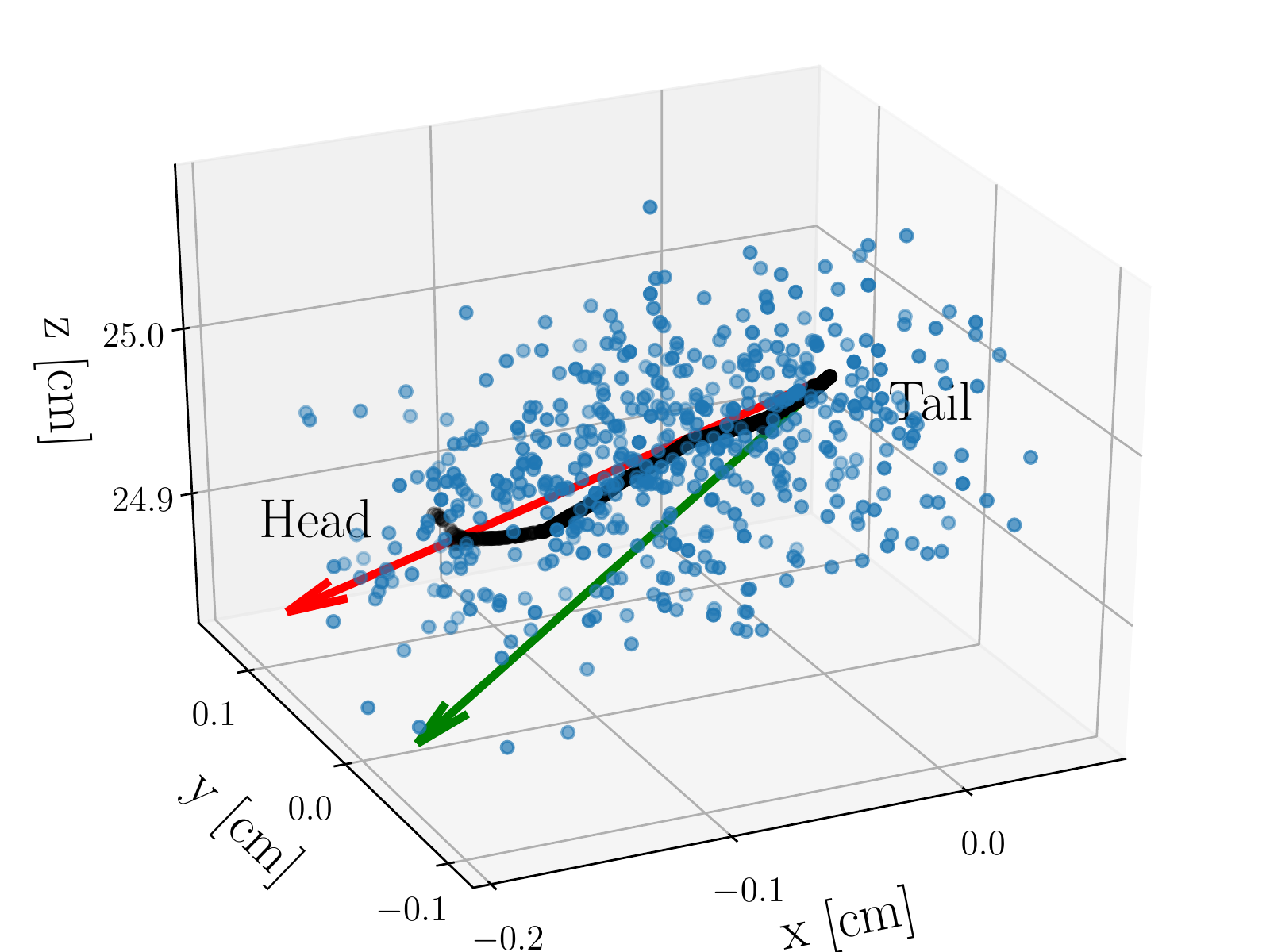}
\end{minipage}
\begin{minipage}{.33\textwidth}
  \centering
  \textbf{Electron recoil}
  \includegraphics[trim = 5mm 0mm 12mm 0mm, clip,width=0.99\textwidth]{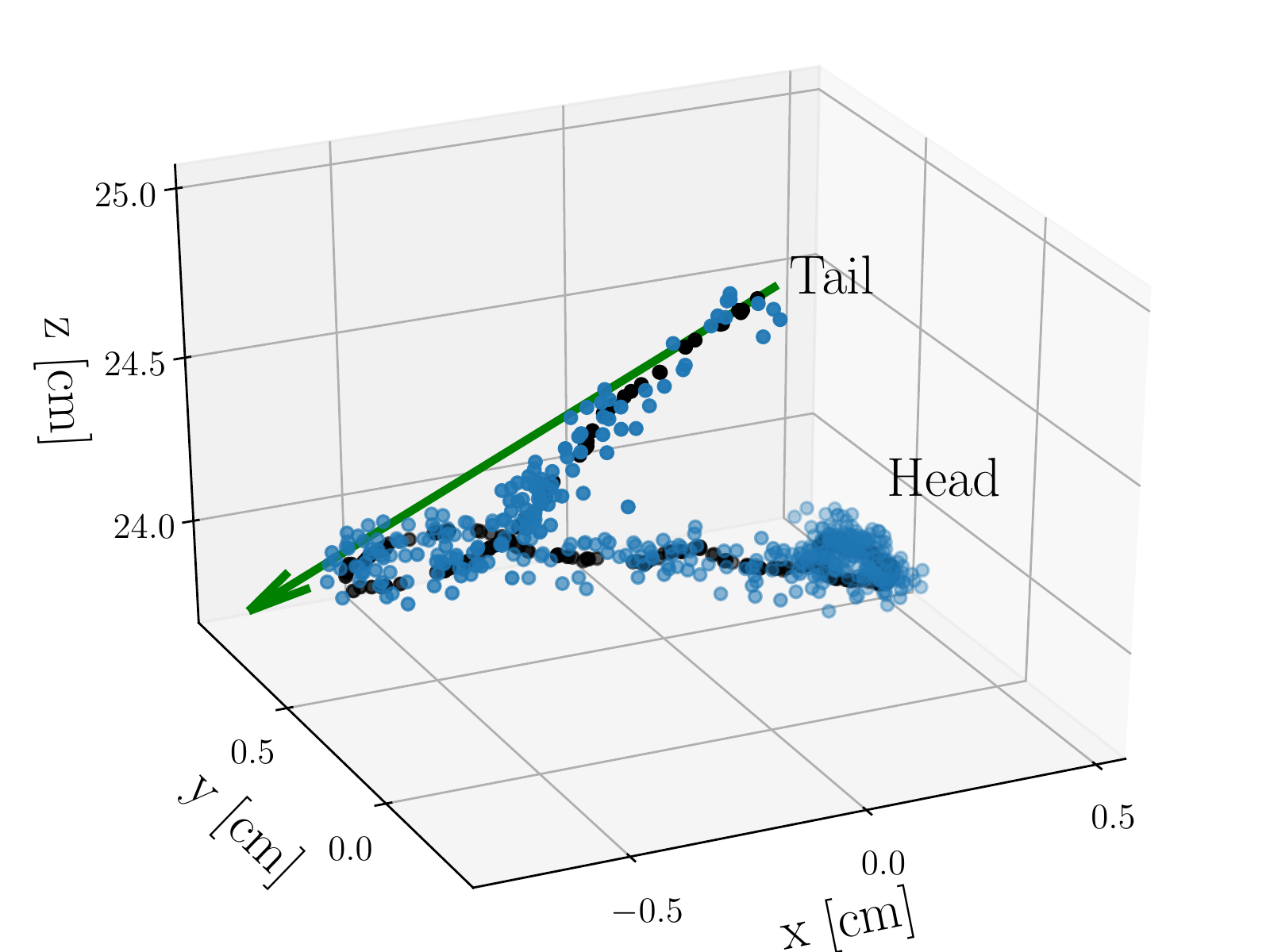}
\end{minipage}
\caption{Simulation illustrating true and reconstructed recoil directions. Black points shows ionized electrons created by a 41~\keVr fluorine recoil (left), a 25~\keVr helium recoil (middle), and a 20~keV electron recoil (right) in atmospheric pressure He:SF$_6$ gas. Note that the electron recoil is about one order of magnitude longer than the two nuclear recoils. Due to ionization quenching, the ionization is nearly the same in these three events, despite the different recoil energies. Blue points show the same ionized electrons after a diffusion of $\sigma_{x,y,z} = 393\,\upmu{\rm m}$, typical for a gas TPC. The reconstructed nuclear recoil direction, red, clearly differs from the true recoil direction, shown in green. The curved recoil trajectory and the diffuse nature of the charge cloud both contribute to this measurement error. In the case of fluorine, the short recoil length and secondary recoils make the direction measurement particularly hard. For electron recoils, a straight-line track fit is clearly not applicable --- a dedicated curled-track fitter would be required.}
\label{fig:recoil}
\end{minipage}}
\end{figure}

Having reviewed the motivation for directional recoil detectors, we now consider how directional information is created by the nuclear and electronic recoil process. We then consider broad classes of technologies as well as specific detectors that can extract this information.


\subsection{Ionization distributions from recoils}\label{sec:recoilphysics}
{\bf Figure~\ref{fig:recoil}} shows the typical primary ionization trails created by 20~\keVee recoils of different types. The energy loss processes of recoils in this low energy regime were first described by Lindhard {\it et al.}~\cite{osti_4701226}; Reference~\cite{Sciolla_2009} provides a thorough review of this physics. At these low energies the energy loss, $\mathrm{d}E/\mathrm{d}x$, is described by the stopping side of the Bragg peak, i.e. decreases as the ion slows down.\begin{marginnote}
\entry{Bragg peak}{the energy loss of a fast, charged particle moving through a medium rises to a maximum value --- the Bragg peak --- before falling steeply as it comes to a stop.}
\end{marginnote}
Energy loss of fast ions occurs mostly by exciting and ionizing atoms along their path (electronic $\mathrm{d}E/\mathrm{d}x$), but as they slow down the dominant process becomes elastic nuclear scattering (nuclear $\mathrm{d}E/\mathrm{d}x$).
The ratio of electronic to total $\mathrm{d}E/\mathrm{d}x$---referred to as the nuclear quenching factor---is a function of recoil energy, the type of recoiling atom, and the composition of the medium. Generally, the transition to where nuclear scattering becomes the dominant channel for energy loss occurs at much higher energies for heavier nuclei.
This complexity in the energy loss of recoiling nuclei is important for DM experiments in general, but is even more crucial for directional experiments.

In directional experiments, the negative slope of $\mathrm{d}E/\mathrm{d}x$ along the trajectory past the Bragg peak provides a measurable attribute to determine the vector head/tail of the recoil track. Naively, head/tail sensitivity should improve when taking quenching into account since most directional technologies rely on measuring ionization along tracks. In practice this is not the case, however, since the nuclear energy loss results in secondary recoils, which also lose energy in both nuclear and electronic channels, resulting in a cascade. The end result is that the energy loss of the primary nuclear recoil is diffused into the surrounding region lateral to its direction. This dilutes the head/tail signature and produces a shortened projection of the track along the detection planes, both from multiple scattering and the diversion of energy lateral to the main track. This was explicitly shown in References~\cite{Majewski:2009an,Deaconu:2017vam}, where simulations based on Stopping and Range of Ions in Solids (SRIM)~\cite{SRIM} were used to study the limitations of head/tail reconstruction in directional DM searches. 

Being a statistical process, multiple scattering results in fluctuations in the energy loss and range of the ion as well as deviations in the recoil's path from its initial direction --- generically referred to as ``straggling''. 
The deviations from a straight path due to multiple scattering can be described statistically  by defining an angular resolution. In general, this quantity has a number of contributions, such as diffusion and the resolution of the detector readouts, however, the physics of energy loss described here poses a fundamental limit. There are multiple conventions for angular resolution, we propose one suitable for comparing directional detectors in Section~\ref{sec:angularperformance}.

Another consequence of nuclear energy loss is how it affects discrimination between electron and nuclear recoils. The classic method relies on the differences in $\mathrm{d}E/\mathrm{d}x$ between these particles. Discrimination via this method must break down at low energies due to quenching, even if the effects of diffusion on short tracks could be neglected; the ionization $\mathrm{d}E/\mathrm{d}x$ of nuclear recoils and electrons converges if quenching grows at low energies, as it is expected to. This is exacerbated by the effective $\mathrm{d}E/\mathrm{d}x$ of electrons, which appears to grow as they slow down due to the rapid increase in scattering causing their tracks to curl up at the end of their trajectories. This is seen in both the simulated ({\bf Figure~\ref{fig:recoil}}) and measured electron tracks. This results in the slope of the $\mathrm{d}E/\mathrm{d}x$ having the opposite sign  
along the tracks of low energy nuclear recoils relative to those for electrons. This is critical for detecting the Migdal effect with directional detectors, as described in Section~\ref{sec:migdal}.



\subsection{Recoil imaging versus indirect direction measurement}\label{sec:recoilimaging_vs_indirect}

\begin{figure}[t]
  \checkoddpage
  \edef\side{\ifoddpage l\else r\fi}%
  \makebox[\textwidth][\side]{%
    \begin{minipage}[t]{1.0\textwidth}
      \centering
	\includegraphics[width=1.0\textwidth]{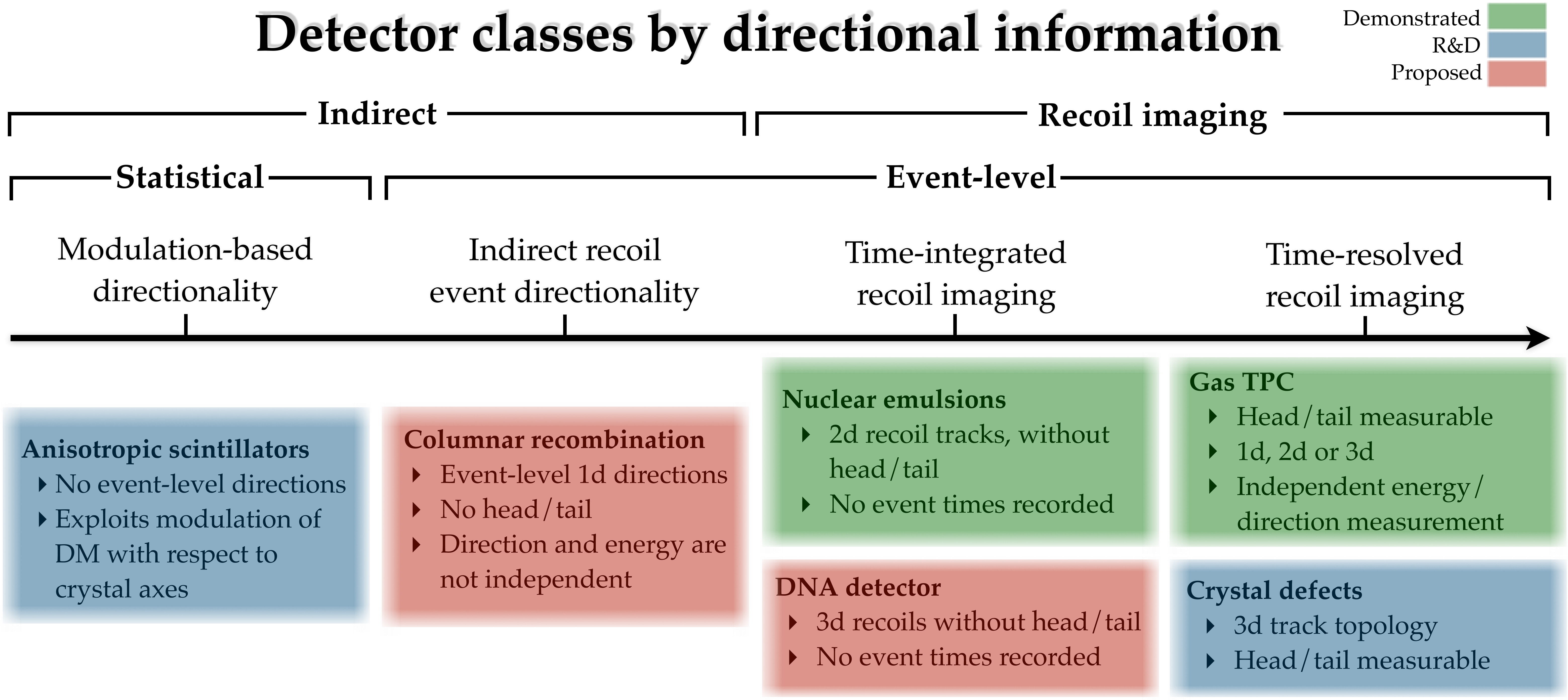}
	\caption{Categorization of different directional detection strategies ordered left to right from the lowest to highest degree of directional information they provide. We also color-code each strategy according to its technological readiness.}
	\label{fig:DetectorClassTable}
	\end{minipage}
}
\end{figure}

There are two broad strategies for detecting recoil directions: directly imaging the recoil track, and indirect methods. We will first describe the primary reasoning behind this distinction, before listing some specific examples of proposed or demonstrated technologies that fall into this categorization. These are also listed in {\bf Figure~\ref{fig:DetectorClassTable}}. We can refine this categorization even further by whether the detector can sense recoil directions at the event level, or with only statistical distributions of events, but we will describe this in more detail when we discuss performance in Section~\ref{sec:performance}.

Recoil imaging entails directly observing one or more components of the recoil trajectory. Referring again to {\bf Figure~\ref{fig:recoil}}, we see that such a capability implies two detector requirements. First, the detector readout segmentation must be smaller than the recoil length of interest, so that multiple space points along the track are obtained. Second, any potential diffusion of the recoil trajectory information must also be small compared to the recoil length, so that the trajectory is not washed out. Additionally for electron recoils, good sensitivity to low densities of energy deposition is also required.\begin{marginnote}
\entry{Micropattern gaseous detectors (MPGDs)}{gas avalanche devices, such as the gas electron multiplier (GEM), with 100-$\upmu$m-level feature size, enabled by modern photo-lithographic fabrication techniques.}
\end{marginnote}
These requirements for achieving recoil imaging are satisfied in gas and solid targets, but not in liquid. In low-density gas TPCs, keV-scale nuclear recoil lengths are $\mathcal{O}$(mm), while the segmentation of modern readouts, such as micro pattern gaseous detectors (MPGDs)~\cite{Oed:1988jh,Sauli:1999nt}, and diffusion are both $\mathcal{O}(100~\upmu$m). In condensed matter, recoils are about three orders of magnitude shorter, while diffusion of ionization is comparatively large. Nevertheless, such detectors can detect the topology and $\mathrm{d}E/\mathrm{d}x$ of higher-energy recoils and utilize this for particle identification. This is the case in DAMIC~\cite{Aguilar-Arevalo:2015lvd}, which can detect ionization energies in silicon as low as $50$~eV in $25~\upmu{\rm m} \times 25~\upmu{\rm m}$ pixels, but is diffusion limited for low energy recoils whose physical track lengths are shorter than $15~\upmu{\rm m}$. Drifting of ionization for near-real-time track imaging thus appears feasible only in gas. In solid targets, because the atoms do not move, there is also the option of performing ultra-high-resolution recoil imaging via other means, but not in real time. Nuclear emulsions~\cite{Gorbunov:2020wfj}, are an example of this strategy.

Given the target mass advantage of condensed matter over gas, and the technological challenges of recoil imaging in the former, it is highly desirable to seek entirely different strategies for obtaining directional recoil information. In contrast to directionality via recoil imaging, we can define \emph{indirectly directional} detectors as those which utilize a variable which has a recoil-direction-dependent response. Anisotropic scintillators for example~\cite{Belli:2020hit}, have a light yield for recoils that depends on the relative orientation of the crystal axis and the recoil. For a single event, there is thus an ambiguity between energy and angle. Nevertheless, from a large data set, variations in angle can be inferred via a sidereal daily modulation in the distributions of detector responses. 

\subsection{Recoil Imaging Detectors}\label{sec:recoil_imaging_detectors}

\begin{figure}[t]
  \checkoddpage
  \edef\side{\ifoddpage l\else r\fi}%
  \makebox[\textwidth][\side]{%
    \begin{minipage}[t]{1.2\textwidth}
      \centering
      \includegraphics[width=0.9\textwidth]{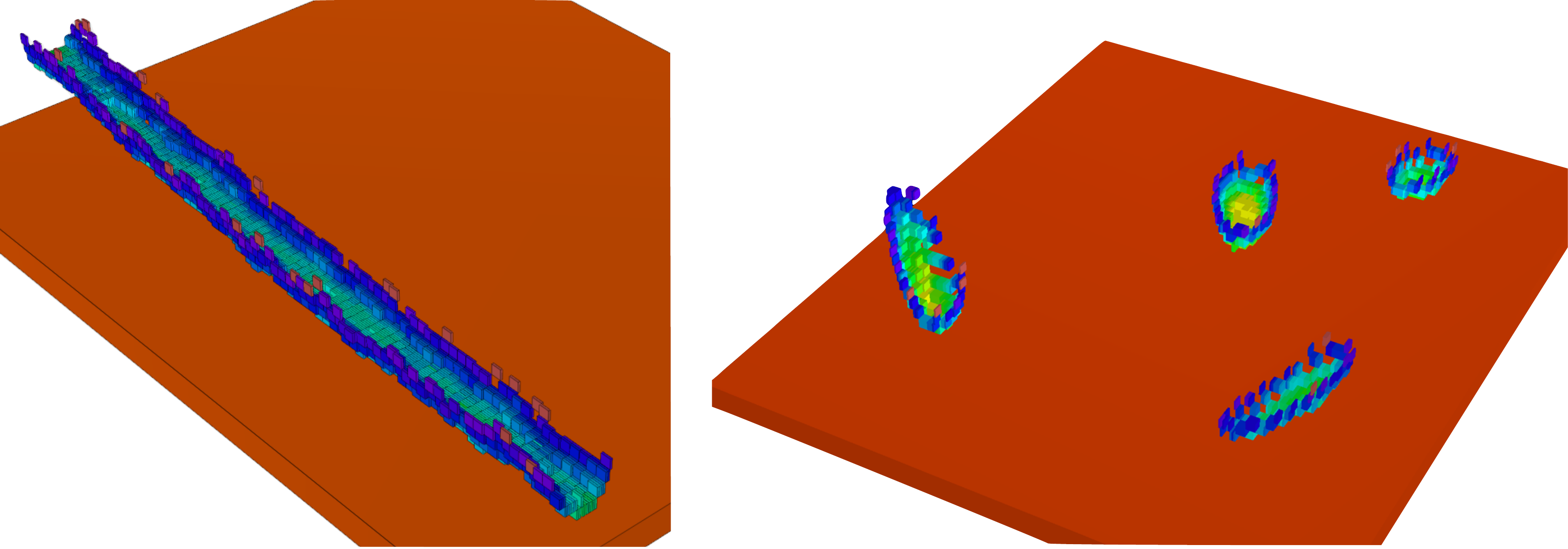}
          \caption{Example of 3d ionization distributions measured with a high definition time projection chamber (HD TPC). Left: an alpha particle track. Right: four superimposed (likely helium) recoil tracks, induced with a neutron source. Each 3d box shown indicates the amount of ionization recorded in a 2d \SI{250}{\micro\meter} $\times$ \SI{50}{\micro\meter} pixel of the TPC readout plane. The vertical coordinate is assigned using the arrival time of the charge on the readout plane. Images taken from Reference~\cite{Jaegle:2019jpx}.}\label{fig:tpc_event}
    \end{minipage}%
  }%
\end{figure}

\begin{marginnote}[]
\entry{Time projection chamber (TPC)}{particle detector capable of imaging ionization in three dimensions by drifting it onto a readout plane, where two-dimensional projections are recorded at high rate~\cite{NygrenTPC}.}
\end{marginnote}

\vspace{1em} \noindent {\bf Gas TPCs.}
The most mature technology used for directional DM searches is the gaseous TPC. This technology provides tremendous flexibility that enables a broad range of operating pressures, $\sim$0.1--1 bar, and the ability to tailor the experiment for the DM parameter space of interest by tuning the gas mixture. At low pressures, the low energy nuclear recoil tracks expected from DM interactions can reach lengths of a few mm: long enough to be resolved and have their directions reconstructed. Gas mixtures can be chosen to include light or heavy targets as needed to optimize for the DM mass range of interest. Gases can also be chosen to include DM target nuclei with large nuclear spin (e.g.~fluorine) or large numbers of nucleons (e.g.~xenon) to enhance sensitivity to SD or SI interactions respectively.

TPC readouts (discussed below) can provide 1d, 2d or 3d track reconstruction with a granularity of $\sim$200~$\upmu$m or better for each track component. For the lateral track components ($x$-$y$) this became possible with the advent of MPGD technologies, whereas for the $z-$component---along the drift direction---the standard approach of using pulse-shape timing together with the drift velocity is used. {\bf Figure~\ref{fig:tpc_event}} shows examples of such TPC measurements. In that case, the ionization was imaged in 2d with a high definition pixel chip readout, at a rate of 80~MHz.

A number of advances over the past decade have improved the sensitivity of TPCs for directional DM searches. One important discovery early on~\cite{Martoff:2000wi} was the effect of adding electronegative components to the gas mixture. This enables negative ion drift (NID), which results in very low diffusion in the thermal regime and a factor $10^3$ slower drift speeds compared to electron drift. The former leads to better ability to resolve the short low-energy ionization tracks even over long drift lengths; lengthening this dimension is also a more cost effective path to scaling up the TPC volume. The slow NID speeds provide $<$100 $\upmu$m pixelization of the track along the drift direction with simple off-the-shelf electronics, resulting in exquisite resolution of this component at low cost.

\begin{marginnote}[]
\entry{Fiducialization}{rejection of background events, typically originating from radioactive contamination of detector surfaces, by reconstructing the absolute spatial position of such events and vetoing a specific spatial region.}
\end{marginnote}
Another critical advance has been the discovery of several methods to fiducialize events in the $z$-direction. This proved challenging due to the lack of a ``t-zero'' reference time for when the event occurred in the TPC, which can be used together with the drift speed to reconstruct its $z$ location.
One method followed the serendipitous discovery of secondary negative ion species in NID gas mixtures that, due to their different drift speeds relative to the primary's, provide an event-by-event reconstruction of $z$ to sub-cm precision. These ``minority carriers'' were discovered in both CS$_2$:O$_2$ \cite{Snowden-Ifft:2014taa} and SF$_6$ \cite{Phan:2016veo} and, with the former, led to a transformation in the field by demonstrating zero-background limits in directional DM searches \cite{Battat:2014van, Battat:2016xxe}. 
A second method that determines the drift distance $z$ by measuring the transverse diffusion along the track, has also been demonstrated~\cite{Lewis:2014poa}. This technique should work in either electron or NID gases, but requires a detector readout segment size smaller than the typical diffusion scale, i.e. it requires a HD TPC. 

With all these advances in gas TPCs the biggest challenge for directional DM searches remains the low density target. Over the past decade the best SD limits set by directional experiments have been surpassed by nondirectional ones by many orders of magnitude. To remain competitive, while maintaining all of the desired features (Section~\ref{sec:performance}) required to detect directionality on an event-level basis, the obvious path to scale-up was to increase the detector volumes by many orders of magnitude. This approach has been reassessed recently, however, due to the looming neutrino floor and lack of hints consistent with the standard WIMP paradigm, which has motivated a search for new classes of DM candidates. Many of these lie in the sub-10 GeV mass range and fall under the umbrella of ``light DM''.
This is an area in which directional experiments using lighter target nuclei such as helium and hydrogen, and possibly also exploiting the electron recoil signature, could become competitive, even with target volumes that could be reached with current technologies. The major requirements for directional light DM searches, where the directional thresholds need to be as low as possible, is spatial resolution. As there are now many options for readouts that satisfy this requirement (see below), the choice comes down to cost, scalability, and other considerations such as backgrounds. 

\vspace{1em}  \noindent {\bf Dark matter TPC projects and readouts.}
We provide a brief historical overview of gas TPC projects, which have enjoyed the most interest. A more comprehensive review of directional detection technologies and relevant TPC readouts can be found in Reference~\cite{Battat:2016pap}.

The first directional DM detector was a low pressure $\sim$100 liter gas TPC with an optically read out parallel plate avalanche chamber (PPAC)~\cite{Buckland:1994gc}. Two gas mixtures were used, 20 Torr CH$_4$ (H target) and 50 Torr of P-10 (90:10 Ar:CH$_4$, Ar as the target) with a $\sim$7\% additive of triethylamine (TEA) vapor used in both to enhance the photon yield.  The PPAC gave very high, 10$^5$--10$^6$ gas gains and light yield peaking in the UV. The two-dimensional track images in the PPAC plane were imaged with a multi-stage optical system that consisted of a UV grade lens, an image intensifier, a second lens, and a CCD camera. The TPC itself was placed inside a superconducting magnet with a 4.5~kG $B$-field parallel to the drift $E$-field. The $B$-field served to reduce transverse diffusion to $<$1~mm over 1~m of drift, and it also deflected electron tracks, producing helical/spherical shapes when projected on the image plane. The resulting topological features were used to reach a 99.8\% rejection of gamma/electron events with a 75\% nuclear recoil efficiency, above an energy threshold of about 6~\keVee. Although the authors of Reference~\cite{Buckland:1994gc} demonstrated many important features required for directionality with this detector, it was never deployed underground. 

The Directional Recoil Identification from Tracks (DRIFT) experiment~\cite{Battat:2016xxe} was the first directional DM experiment to take underground data and continued to do so with several generations of detectors over a decade-long program. The DRIFT detector was based on a m$^3$ TPC divided by a central cathode into two halves, each read out with a Multi Wire Proportional Chamber (MWPC). Signals from the MWPC anode wires (2 mm pitch) and their pulse-shape timing provided two components of ionization tracks, $\Delta x$ and $\Delta z$, respectively. DRIFT pioneered the use of NID with CS$_2$ gas mixtures, which provided thermal diffusion and slow drift. These features enabled 2d tracking with head/tail reconstruction in 1d ($z$)~\cite{ref:DRIFT_APP2007,Battat:2016xaw} and a gamma/electron rejection factor of $\sim 10^{-7}$ for energies between 18--150 \keVee \cite{Battat:2016xxe}. DRIFT developed novel techniques to reduce and mitigate against backgrounds from radon and its progeny~\cite{Battat:2014oqa, Battat:2015rna}, which allowed full fiducialization of the detector volume~\cite{Battat:2014van}. Culminating this effort, DRIFT set a series of zero-background DM limits in CS$_2$:CF$_4$:O$_2$ gas mixtures with underground data taken at the Boulby Mine in the UK~\cite{Battat:2016xxe}. Their most competitive limit, set using 54.7 live-days, was $\sigma^{\rm SD}_p < 2.8 \times 10^{-37}$~cm$^2$ at $m_\chi = 100$~GeV. A combination of the coarse granularity and low S/N of the MWPCs, a 
low target mass (34 g of fluorine in total) and relatively high energy thresholds (about 35 keVr for fluorine recoils) limited DRIFT's directional and DM sensitivity. 

The advances over the past decade in MPGDs and commercially available scientific-grade CCD/CMOS sensors  led to a number of new TPC-based directional experiments. The Dark Matter Time Projection Chamber (DMTPC)  collaboration built a series of prototypes with CCD-based optical readouts that imaged 2d tracks at the surface of a mesh avalanche stage~\cite{Deaconu:2015vbk}. They set a limit of $\sigma^{\rm SD}_p < 2\times10^{-33}$~cm$^2$ at $m_\chi = 115$~GeV, from a surface run in pure CF$_4$ gas~\cite{Ahlen:2010ub}. A 1~m$^3$ detector was constructed, but to our knowledge not deployed~\cite{Leyton:2016nit}. The CYGNO collaboration employs thin GEMs read out optically with CMOS cameras. They plan to augment the 2d optically imaged tracks using pulse-shape timing with a PMT to measure the third dimension. With several prototypes they have performed R\&D using 1 bar He:CF$_4$ gas mixtures that are being optimized for light DM and solar neutrinos~\cite{Baracchini:2020btb}. Their short-term program involves deploying a $\sim$m$^3$-scale demonstrator in the Gran Sasso National Laboratory, with scale-ups to $\sim$10 m$^3$ in the future. 

Electronic readouts using Micromegas, GEMs and other novel MPGDs for gas amplification combined with strips or pixels are also being used both for R\&D and in underground experiments. The NEw generation WIMP-search with Advanced Gaseous tracking device Experiment (NEWAGE) collaboration has deployed several generations of TPC detectors in the Kamioka Mine. Their technology is based on a micro pixel chamber ($\upmu$-PIC) combined with GEMs and strip readouts that provides them with vector 3d tracking. From measurements in 76 Torr CF$_4$ they 
report a an electron/gamma rejection of $\sim 10^{-5}$,
a correct head/tail sense determination of 53.4\% and an angular resolution of 36$\pm$4$^\circ$, all for 50--100 \keVee~\cite{Yakabe:2020rua}.
Although limited by radon backgrounds, they have used directionality to set several limits, with the latest reported at $\sigma^{\rm SD}_p <$ 4.3$\times$10$^{-34}$ at $m_\chi = 150$~GeV~\cite{Yakabe:2020rua}. The MIcro-tpc MAtrix of Chambers (MIMAC) experiment uses a Micromegas pixel readout TPC with a special gas mixture  (70:28:2 CF$_4$:CHF$_3$:C$_4$H$_{10}$) tuned for SD sensitivity and other properties that allow full 3d tracking \cite{Santos:2011kf}. MIMAC is located in the Modane underground laboratory in France, but DM limits have not been published as of yet. 

The Directional Dark Matter Detector (D$^3$) project, an R\&D collaboration between Lawrence Berkeley National Laboratory and the University of Hawaii, has constructed small TPC prototype detectors with high-definition pixel charge readout based on 
application-specific integrated circuit (ASIC) chips. Eight of the latest generation detectors, also known as the ``BEAST TPCs''~\cite{Jaegle:2019jpx} were deployed for directional neutron background measurements at the SuperKEKB collider, using a He:CO$_2$ target gas mixture. While the target mass was minute, and the detectors running in low-gain neutron mode, a DM limit extending down to as low as 4~GeV was set as a feasibility demonstration~\cite{phdthorpe}. Events from these detectors are shown in {\bf Figure~\ref{fig:tpc_event}}. Since they can efficiently detect single electrons at higher gain settings, excellent low-mass DM sensitivity is expected. While considerable cost and effort is required to scale up such detectors to competitive masses, larger-scale pixel based readout planes are already being fabricated and tested for tracking detectors in future colliders. This is an R\&D synergy that could prove useful for the field. In fact, GridPix detectors~\cite{Ligtenberg:2020ofy}, based on pixel ASICs that are directly combined with a gas amplification MPGD structure, have already demonstrated exquisite imaging of nuclear recoils~\cite{Kohli:2017qzo}, with even finer spatial segmentation than in {\bf Figure~\ref{fig:tpc_event}}.

Given the abundance of available TPC charge readout technologies, it is not straightforward to determine the best strategy for a large-scale detector. The recent \Cygnus design study~\cite{Vahsen:2020pzb} is the first attempt at such a technology comparison, and suggested that x/y strips with $\mathcal{O}(100\,\upmu{\rm m})$ segmentation provide the best cost/performance tradeoff. An optimized strip readout should enable HD charge readout near the resolution obtained with pixel ASICs, but at substantially reduced cost and complexity. Based on this, two (40 liter and 1000 liter) ``\Cygnus HD demonstrator'' detectors, utilizing CERN strip Micromegas readout and CERN SRS DAQ systems, are now under construction~\cite{vahsen_aps_2020}.

\vspace{1em}  \noindent {\bf Nuclear Emulsions.}
The low densities of gas-based experiments is the primary factor working against them being the obvious strategy for directional recoil detection. Technologies that can image recoil tracks in high density materials are therefore highly motivated. However, as we discussed earlier, an increase in density must always be matched by an increase in spatial resolution due to the rapidly shrinking track lengths. One long-standing technology that is both high density and could permit the necessarily high spatial resolution, are nuclear emulsions. Nuclear emulsions consist of photographic plates with some dispersal of smaller crystals or grains. The nuclear emulsion most well developed for low energy nuclear recoils consists of a polymer layer dispersed with silver halide (AgBr) crystals. The crystal grains would seed nm-scale silver clusters in response to a track left by a recoil. After a suitable exposure time has elapsed, the emulsions must then be developed, during which 2d projections of recoil tracks can be identified and measured with an optical or x-ray microscope. The Nuclear Emulsions for WIMP Search (NEWSdm) collaboration~\cite{Gorbunov:2020wfj} is pursuing this idea with an automated optical scanning system. They employ a Nano Imaging Tracker~\cite{FineGrained} which can measure the positions of single grains with an accuracy of 10~nm. Nuclear emulsions are capable of 2d (and potentially 3d) event-level recoil imaging, however, the presence of any head/tail signature is unclear. Event time assignment is not possible with this technology.

\vspace{1em}  \noindent {\bf Crystal defects.}
Another potential solid state directional detector involves the imaging of crystal damage, in particular in diamond. Nitrogen vacancy (NV) centers in diamond are defects consisting of nitrogen impurities neighboring a vacancy in the crystal lattice. The defects are highly sensitive to electromagnetic fields, and to the local crystal strain. Spectroscopically measuring NV centers in diamond has been suggested as a potentially promising way to image nm-scale crystal damage that could be left by a recoil~\cite{Rajendran:2017ynw}. This technology would benefit from the high densities of crystals, and may allow fully three-dimensional recoil imaging, with even a plausible head/tail signature. A recent study~\cite{Marshall:2020azl} expanded upon this idea and suggested a modular design strategy that, if realized, would bring the available quantity of directional information up to the same level as gas TPCs while enjoying a high inherent mass. 

\vspace{1em}  \noindent {\bf 2d materials.}
When imaging recoils in higher density targets, the ultimate recoil trajectory will be less correlated with the true initial recoil direction due to the high rate of interactions in the medium. A way to sidestep this problem may be to exploit materials in which the medium itself is confined to two dimensions. Such 2d targets could be fabricated from semiconductor materials in which the excitation energy is on the order of 1~eV, allowing even light DM particles to generate measurable electronic events. 
In the set up envisaged in Reference~\cite{Hochberg:2016ntt}, a 2d detector is comprised of an array of pixels containing back-to-back layers of graphene and a substrate, all placed inside an applied electric field to transport excited valence electrons to a calorimeter. This configuration can automatically exploit modulation-based directionality using the contrast in event numbers in the upper and lower layers of pixels. However, it is proposed that if one could monitor the conductivity of the graphene pixel arrays fast enough relative to the drift time of the electron, one could reconstruct the precise locations of pixels at which the electron interacts and the time-of-flight. As long as the experiment were in a high enough vacuum, in principle, this information can be manipulated to reconstruct the full three-dimensional recoil vector. No experimental implementation of this idea has yet appeared, however the relic neutrino experiment PTOLEMY~\cite{Betti:2019ouf} proposes to use tritiated graphene, so this effort could be complementary.

\vspace{1em}  \noindent {\bf DNA strands.}
A novel re-imagining of a recoil imaging detector proposed in Reference~\cite{Drukier:2012hj} makes use of a forest of DNA or RNA strands hung vertically from a nanometer-thick gold foil. An incoming particle would collide with, and expel, a gold atom from the foil, and the recoil would then travel through the DNA forest, severing several strands. The strands would be sequenced precisely with base pairs that encode their $(x,y,z)$ positions in the detector volume. Using a well-established biotechnology known as polymerase chain reaction, it would be possible to amplify the severed strands once collected, reconstruct the positions of each strand break, and therefore the coordinates of each severing event to nm precision. This represents an entirely different method of imaging the nuclear recoil axis that negates the effects of diffusion. However, without any published demonstrations of this idea it is not clear whether it is practical, and to what extent head/tail and timing information could be recorded.

\subsection{Indirect Directionality}
\vspace{1em}  \noindent {\bf Anisotropic scintillators.}
Solid scintillators (e.g.~NaI and CsI) are commonly used in particle detection, and specifically in DM detection. Some scintillators, such as ZnWO$_4$, have been shown to exhibit a response that depends on the recoil direction relative to the crystal axes. In principle, this scintillation anisotropy can be used to infer the nuclear recoil track direction without direct reconstruction of the track geometry. Several groups have explored the possibility of using anisotropic scintillators for a DM search~\cite{Shimizu:2002ik,Cappella:2013rua,Sekiya:2003wf}. The anisotropy of ZnWO$_4$ in particular has recently been confirmed via measurement with a neutron gun for the ADAMO project~\cite{Belli:2020hit}, but only for energies higher than 70~\keVee. However, even if a strong anisotropy were discovered at lower energies, this form of directionality would be indirect, and achieving event-level directionality would likely be impossible. A DM search using anisotropic scintillators would have to exploit the daily modulation in events as the DM wind rotated with respect to the crystal axes.

\vspace{1em}  \noindent {\bf Columnar Recombination.}
An indirect measure of recoil directionality called columnar recombination may be present in high pressure gaseous xenon or liquid argon (LAr) experiments. The effect appears when there is an asymmetry in the way an ionisation cloud generated by a recoil event behaves depending on its orientation with respect to an applied electric field. When a primary ionisation cloud drifts in the field, some of the ions and electrons will recombine. The amount of ionisation or scintillation that is ultimately detected from the event may then depend on the axial angle between the recoil track and the electric field. In principle, parallel tracks would produce more scintillation, and perpendicular tracks more ionisation. 
This effect was suggested to be of potential use for DM searches by Nygren~\cite{Nygren:2013nda}. Subsequently, it has been investigated experimentally using nuclear recoils in LAr by SCENE~\cite{Cao:2014gns}. However, the measured directional asymmetry in the scintillation yield of neutron-induced recoils was small and only statistically significant for their 57 keV beam. Nevertheless, since planned LAr experiments such as DarkSide-20k~\cite{Aalseth:2017fik} and Argo~\cite{Sanfilippo:2019amq}, are anticipated to reach the neutrino floor in the next few decades, a directional signal in this kind of experiment would be highly sought after. Unfortunately, since columnar recombination only provides information about one track dimension, does not have a head/tail signature, and must be obtained through a combination of two recoil energy observables; it ultimately appears to be somewhat insufficient for the discovery of DM~\cite{OHare:2020lva}. 

\section{DIRECTIONAL DETECTOR PERFORMANCE}\label{sec:performance}

The ideal DM detector, whether directional or not, will have high target mass and low energy threshold, so as to maximize the probability of observing DM signals. To avoid becoming background limited, a low-background (underground) environment, highly radiopure components, and good background rejection capabilities are also required. An ideal {\it directional} DM detector is subject to the same requirements, but would in addition need to measure each recoil’s 3d vector direction, energy, topology, and time of occurrence. Below, we use back-of-the-envelope arguments and simulations to make quantitative performance requirements for each of these observables. Since we, perhaps subjectively, believe that gas TPCs are the closest to providing the required performance, we will use these detectors to illustrate state-of-the art performance and to suggest directions for future work.


Directional recoil detection is still a young and growing field, and it is not straightforward to compare angular performance across different detector types. In much of the literature to date~\cite{Ahlen:2009ev, Mayet:2016zxu, Battat:2016pap}, directional detectors have been classified and simulated as head/tail sensitive or not, and as 1d, 2d, or 3d, depending on how many projections of nuclear recoils are detected. To give a more holistic view of the field we introduce a new classification, extending the concepts introduced in Reference~\cite{Vahsen:2020pzb}. The complete scheme is depicted in {\bf Figure~\ref{fig:DetectorClassTable}}. We already introduced the distinction between recoil imaging and indirect detection of the recoil direction. In comparing the physics sensitivity of detectors, a second, slightly different classification is useful. This classification is based on information content: we can classify detectors by their ability to gain directionality at the event-level, or statistically via modulating signals.

The first category, event-level directional detectors, includes most, but not all, proposed directional detectors. These detectors directly reconstruct (or infer some component of) the recoil vector event-by-event. Examples of detector technologies in this category are: gas time projection chambers, nuclear emulsions, crystal defect spectroscopy, and detectors based on 2d graphene targets. Detectors that infer the recoil direction event-by-event by measuring two different physical quantities, for example the energy and one recoil trajectory component (gas TPC with 1d readout), or the ionization and scintillation energy (liquid noble gas TPC with columnar recombination), also belong in this category. 

On the other hand, if directional information is only present at the level of a statistical distribution of recoils, then a discovery or rejection of isotropy can only be performed using the modulation of that recoil distribution. Examples of this category are anisotropic scintillators. However, we note that if a second recoil observable were available for each event, such as energy in a different channel (ionization, heat) the observational degeneracy between energy and angle may be broken. In that case, the detector could obtain directional information at the event-level. 

Both event-level and modulation-based directional detectors can be said to have directional sensitivity, and all such detectors could, in principle, verify the galactic origin of a DM signal to a greater or lesser extent. Event-level directional detectors could exploit both the dipole feature (upper panel of {\bf Figure~\ref{fig:Skymaps}}) and the sidereal modulation (lower panels of {Figure~\ref{fig:Skymaps}}), whereas modulation-based detectors are forced to rely only on the latter. Both methods of directionality have powerful and unique signatures of DM that should not be mimicked by any background or systematic. 

One other important distinction between event-level and modulation-based directional detectors, is that in the latter, direction and energy are not measured independently. In the context of neutrino measurements, this means that if the neutrino source location is known (as in the case of the Sun, a supernova, or a neutrino beam) then independent recoil energy and direction measurements can be combined to calculate the neutrino energy, event-by-event. This powerful capability is not available in modulation-based detectors.

\subsection{Directional performance of event-level directional detectors}\label{sec:angularperformance}
The performance of an event-level directional detector will depend upon how much information about each recoil is measured. An ideal detector would measure the full three-dimensional vector corresponding to the true initial recoil direction, i.e. the direction immediately after the scattering process that produced the recoil. This direction is shown as a green arrow in {\bf Figure~\ref{fig:recoil}}. 

However, the direction of the entire recoil track (red in {\bf Figure~\ref{fig:recoil}}) will generally differ from the direction of the initial recoil due to several effects. First, the recoil does not travel in a straight line due to scattering (also known as straggling for nuclear recoils), so that even a well-measured average recoil direction will deviate from the initial, true recoil direction. Second, detector limitations such as charge diffusion and the finite segmentation of the readout, will further smear the measured recoil vector. Third, and often by design, a particular detector technology may not be able to measure all three components of the recoil vector, or its sign. Importantly, these effects are all energy-dependent, and each leads to worse directionality at lower energies.

Despite these considerable complications, we can still describe the directional performance of any event-level directional detector with two simple and independent quantities: the effective 3d angular resolution and the head/tail recognition efficiency. Because several choices and conventions for these quantities exist, we will first carefully define these, adopting the same conventions we introduced in Reference~\cite{Vahsen:2020pzb}.

\vspace{1em}  \noindent {\bf Angular resolution.} We take this to be the mean difference between the true, initial recoil axis, and the measured recoil axis. The difference is measured by a single angle in a three-dimensional space. This corresponds to the angle between the red and green dashed lines in {\bf Figure~\ref{fig:recoil}}, but does not consider the sign of the two vectors. Because this difference is the angle between two {\it axes} (as opposed to two vectors), it ranges from 0 to only 90 degrees. As a result, the angular resolution ranges from 0 to 1 radian (approximately 57 degrees). Note that the upper limit of 1 radian is the average angle between two randomly chosen axes in 3d, i.e it corresponds to having no (axial) directional sensitivity.

\vspace{1em}  \noindent {\bf Head/tail recognition efficiency.} We define this to be the fraction of events where the vector product of the reconstructed recoil direction and true initial direction is positive. A value of 0.5 corresponds to completely random head/tail assignment by the detector, and 1 is the best possible performance.

\begin{figure}[t]
\centering
  \includegraphics[trim = 0mm 0mm 0mm 0mm, clip, width=1.0\textwidth]{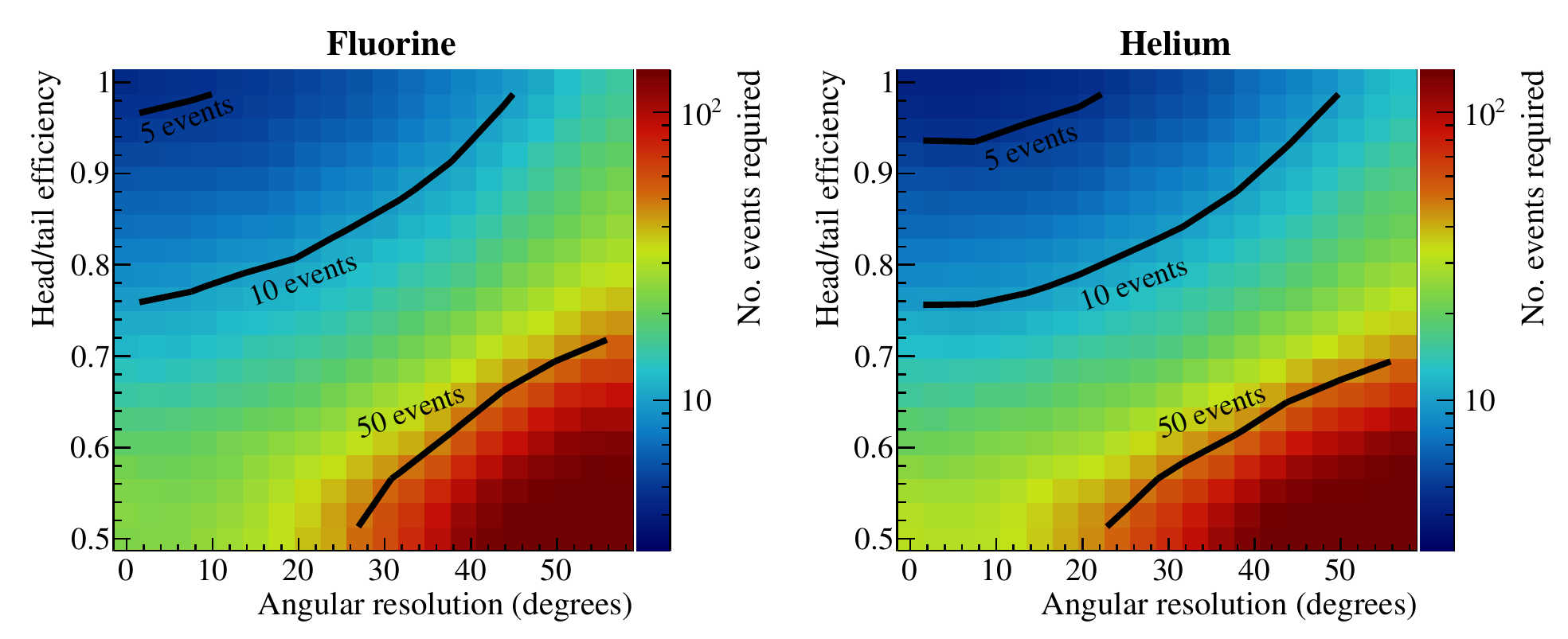}
\caption{Impact of an event-level nuclear recoil detector's directional performance on solar neutrino/DM discrimination. The color scale shows the required number of detected fluorine (left) and helium (right) recoils to exclude a solar neutrino background hypothesis at 90\% C.L., versus angular resolution and head/tail recognition efficiency, as defined in Section~\ref{sec:angularperformance}. This particular simulation assumes $m_\chi = 10$~GeV, a He:SF$_6$ target gas, and an energy threshold of 1~\keVr. The top left of each plot corresponds to an idealized detector, while the bottom right corresponds to no directional sensitivity. The shape of the contours shows that both angular resolution and head/tail efficiency are required for optimal discrimination between WIMPs and solar neutrinos. That said, a detector with only good head/tail recognition (top right corner) performs significantly better than a detector with only good angular resolution (bottom left corner).}
 	\label{fig:required_ang_performance}
\end{figure}

\vspace{1em}  \noindent {\bf Required performance.} The above quantities form a finite parameter space and have the benefit of being robust and easy to measure directly, both in experiment and simulation, without any need for parameterization or fitting. Note that both definitions remain valid and useful even if not all recoil projections are measured. This means that for a given recoil energy, any event-level directional detector can be viewed as one point in the two-dimensional performance parameter space shown in {\bf Figure~\ref{fig:required_ang_performance}}. To show the impact of performance, the figure shows the required number of observed DM-helium or DM-fluorine recoils required to reject a solar neutrino background hypothesis.\footnote{For simplicity, this analysis utilizes only the recoil angle distributions in galactic coordinates integrated over one year. Incorporating event time and recoil energy information would allow for even fewer required events. The results also depend on the statistical testing methodology. We choose to focus on the big picture here, and defer more detailed descriptions to future work.} We see that a good performance target is an angular resolution of 30 degrees or lower and a head/tail efficiency of 80\% or better. This would result in $\lesssim$10 DM recoils needed to exclude a neutrino hypothesis. The shape of the contour in the figure also shows that head/tail recognition is especially important.

\vspace{1em}  \noindent {\bf Performance in practice.} Angular performance is strongly energy dependent. For example, the \Cygnus simulation of optimized gas TPCs~\cite{Vahsen:2020pzb} suggests an angular resolution of 10$^\circ$ and a head/tail efficiency of nearly 100\% is feasible for helium recoils $\gtrsim50$~\keVr. At lower energies, even a highly idealized detector is limited by the primary ionization distribution of the recoils to about 28$^\circ$ resolution and 70\% head/tail efficiency. A realistic gas TPC with diffusion loses most directional sensitivity at 1~\keVr. Since solar neutrinos and $\mathcal{O}({\rm GeV})$ mass WIMPs generate most nuclear recoils below 10~\keVr, the greatest challenge for future detectors will be to extend good directional performance to low energies. 

In designing future detectors, the contribution of TPC readout performance to angular resolution can be reliably predicted, see Equation 5 in Reference~\cite{Vahsen:2015oya}. For mm-length nuclear recoils, this leads to the requirement of highly segmented detectors, with feature size $\mathcal{O} (100\,\upmu {\rm m})$, and low diffusion. The contribution from the spatial shape of the primary ionization distribution, especially below 10 \keVr has, however, large uncertainties, and the same is true for the head/tail efficiency. Because these directly affect the designs of future detectors, it is imperative for the field to validate the commonly used simulation tools at the lowest energies. Validation work using helium nuclei for energies above 50~\keVee and carbon and fluorine above 10~\keVee can be found in Reference~\cite{Deaconu:2017vam}. Fluorine recoil measurements going down to 6~\keVr, can be found in Reference~\cite{Tao:2019wfh}. For progress in this direction, recoil imaging detectors with low pressure, high definition (HD) readouts and minimal diffusion are required.

\subsection{Directional performance of modulation-based directional detectors}
To date, no demonstration has been made of modulation-based directionality at recoil energies relevant for a DM search. However, proposed modulation-based detectors have a natural upper hand for achieving high exposures in that they are based on liquid or solid targets. So it is interesting to compare the required DM event numbers for discovery to see which strategy is optimal. The difficulty is that the width of the expected DM recoil energy spectrum, $\sigma_E$, is quite broad. Therefore, many events, and an asymmetric detector response that is significant compared to this width, are required to detect a DM signal with this strategy. We are not aware of quantitative studies of this in the literature, so we perform a back-of-the envelope calculation, assuming Gaussian statistics. We assume the detector has a direction-dependent energy response, and that all detector data are grouped into two bins based on the sidereal time. We assume that the time at which the bins are divided is such that we expect higher event energies due to the recoil directions in one bin, lower in the other. With a few additional simplifications, the total number of detected events required to reject an isotropic background, in the presence of signal only, at the $s$-$\sigma$ level is
\begin{equation}\label{eq:indirect}
 n = \frac{4\,s^2}{c^2} \,,
\end{equation}
where $c=\Delta E/ \sigma_E$ is the ratio of the difference in mean energies in the two bins to the width of the average energy distribution. This width will be a convolution of the DM recoil spectrum and the detector response. Assuming, for example, $c=1$\%, we find that of order 360,000 events are required for a 3$\sigma$ level exclusion of isotropy. We have described earlier (Equation \ref{eq:reject_isotropy}), that for an idealized event-level directional detector, as few as 10 events are required to reject an isotropic background at the same level of significance.

If we, for example, assume the comparison is between a recoil imaging detector with an SF$_6$ gas target at atmospheric pressure, and an indirectly directional detector with a liquid xenon (LXe) target utilizing S2 only, and further assume SI WIMP-nucleon scattering (benefiting xenon), then we find that we would need a LXe detector of similar physical size as the gas detector to have similar directional sensitivity. This example is hypothetical, as the gas detector would likely need lower pressure, and the parameter $c$ for LXe is unknown. This illustrates, however, that a common argument made against directional gas detectors---size---may not necessarily be valid, given that a large gas TPC operating at room temperature should be easier to construct, and less costly than a cryogenic liquid noble-gas detector of the same size. To decide on the optimal strategy,  directional performance measurements at relevant energies are still needed. We also suggest that a more careful comparison of event-level and modulation-based directionality be carried out in future work.

\subsection{Energy thresholds and tension between target mass and directionality} 
While a lower energy threshold is generally better for DM searches, in the context of directional detection, there are three relevant energy thresholds: the energy threshold for nuclear event detection, for particle ID, and for directionality; ordered from what are typically lowest to highest. For some detectors, including gas TPCs, there can also be a minimum charge density below which events are not detected.

For the near-term goal of distinguishing solar neutrinos and DM, sub 10-\keVr event detection thresholds are required. Detecting events in this energy range is relatively straightforward for modern particle detectors; for example, a high gain gas detector can easily detect a single electron, corresponding to $\sim 25$~eV of ionization energy.

The real challenge is therefore to achieve directionality and rejection of internal backgrounds in the sub-10-keV energy range. For gas-based detectors specifically, one of the biggest design trade-offs is that the directional threshold and the particle ID threshold both improve with lower gas density, while the detector target mass and hence DM sensitivity are reduced. Low density operation---achieved either via pressures of $\mathcal{O}(10-100)$~Torr, or via low-$Z$ gases at atmospheric pressure---is required to achieve adequate directionality. In existing and proposed designs, directionality still tends to gradually roll off below 50~\keVr, but may be useful for DM/neutrino discrimination down to approximately 6~\keVr~\cite{Vahsen:2020pzb}. Further improvements should be investigated. However, particle identification capabilities tend to deteriorate \emph{exponentially} towards lower energies~\cite{Vahsen:2020pzb}, so in the end this is often expected to be the factor that determines the effective energy threshold for analysis.

Finally, if we wish to optimize a future detector for solar neutrinos, as well as DM, we must consider the recoil energy thresholds required to detect the fluxes of interest. We see from Table~\ref{tab:nurates} that around a 5~keV threshold is required detect a reasonable fraction of CNO neutrinos in the electron recoil channel. However, electron tracks have lower charge density than nuclear recoils, making detection more challenging. The ideal detector would therefore have high efficiency for detecting single primary electrons.

\subsection{Particle identification}\label{sec:particleID}
Preliminary simulations of $1000~{\rm m}^3$ gas detectors~\cite{Vahsen:2020pzb} suggest that internal backgrounds will be dominated by electron recoils from Compton scattering of gamma rays from radio-impurities and the detector environment. It may be necessary to reduce such backgrounds via particle ID at the reconstruction level by factors of $\sim 10^4$--$10^5$. Experimental work and simulations (see Reference~\cite{Ghrear:2020pzk} and citations therein) suggest this is feasible with high-definition gas TPCs at 10~\keVee, and perhaps even substantially lower energies. Above this, electron rejection improves exponentially. It is important for the field to demonstrate such electron rejection capability experimentally, and as a function of energy, as this may determine the practical energy threshold of large detectors. We note without providing further details that the same particle ID capabilities can also be used to identify the recoiling nucleus.

\subsection{Energy resolution}
Energy resolution is relevant to directional detection in a number of ways. For an idealized gas TPC, the fractional energy resolution is given by 
\begin{equation}
    \frac{\sigma_E}{E}=\sqrt{n \times (F+f)},\label{eq:energy_resolution}
\end{equation}
where $n$ is the number of ionized electrons, $F$ is the so-called Fano factor which quantifies primary ionization fluctuations, and $f$ is the relative gain variance of the gas amplification device~\cite{thorpe2020}. In practice, this typically leads to a quantitative resolution of order
\begin{equation}
\frac{\sigma_E}{E} = 10\% \sqrt{\frac{5.9~{\rm \keVee}}{E}} \, ,
\end{equation}
which appears sufficient for particle ID in the context of rejecting electrons and retaining nuclear recoils via their difference in specific ionization, $\mathrm{d}E/\mathrm{d}x$. Finite energy resolution will also smear out any indirect or modulation-based directionality, thereby requiring more events for a given observation. This is seen in Equation~\ref{eq:indirect}, where $c$ is inversely proportional to the signal's energy spread, which increases with worse energy resolution. 

In the future, a recoil-imaging detector could be used to reconstruct the energy spectrum of a known neutrino source. We have performed a preliminary study and determined that the gas detector resolution quoted above is sufficient for reconstructing CNO and $^8$B solar neutrino energies. This is particularly promising with electron recoils, which have higher energy and rate than nuclear recoils, at fixed neutrino energy. The higher electron recoil energies also improve the resolution of the reconstructed neutrino energy spectrum. The \Cygnus collaboration~\cite{Vahsen:2020pzb} is now investigating the detector requirements for this measurement.

While the contribution $f$ in Equation~\ref{eq:energy_resolution} can be minimized by optimizing the detector, it could be reduced to zero by utilizing readouts capable of counting individual electrons. In this case the energy resolution would be determined only by primary ionization fluctuations. This was first attempted in Reference \cite{Sorensen:2012qc} with oxygen charge carriers. It may be achievable now with SF$_6$ or CS$_2$+$O_2$ NID and modern, high-speed charge readouts based on MPGDs~\cite{Ligtenberg:2020ofy,Kohli:2017qzo}. This is a natural performance limit to push for and investigate.

\subsection{Event time}
Ideally, a directional detector will not just measure 3d recoil vectors but will also record exact event times. Only if event times are known can a measured directional signal be transformed into galactic coordinates, and the DM dipole signature (displayed in {\bf Figure~\ref{fig:Skymaps}}) be searched for directly. The same is true of the Sun in the context of solar neutrinos. It is natural then to ask how good the time resolution must be to enable these techniques, and what happens when no timing information is present.

In effect, these coordinate transformations rely on timing to deduce the spin angle of the Earth, $\alpha$, which then gives the recoil angle, $\theta$, with respect to the nominal particle source. Any uncertainty in time, $\sigma_t$, would then create an additional uncertainty in these angles,
\begin{equation}
    \sigma_\alpha = \sigma_\theta = \sigma_t \times \frac{360^{\circ}}{24~{\rm hours}} \, .
\end{equation}
To make this timing-induced angular uncertainty smaller than the intrinsic recoil angle resolution of the detector, we would require, say, $\sigma_\alpha < 10^{\circ}$, resulting in a required time resolution of $\sigma_t < 40$~minutes. This requirement is easily met by modern particle detectors. For TPCs, even for the most pessimistic case of NID without absolute position measurement in the drift direction, the maximal drift time uncertainty would be of order 10~ms~\cite{Phan:2016veo}. 

In the limit of no timing information whatsoever, the detector becomes time-integrating. A detailed study of this scenario~\cite{OHare:2017rag} found that time integration does not wash out all directional information, but instead causes an effective exposure penalty of a factor of 2 for the rejection of isotropic backgrounds, but almost an order of magnitude for probing below the neutrino floor. However, this is the best case scenario, when both head/tail and complete 3d recoil tracks are measurable. If this information is not available then much larger penalties in directional sensitivity are incurred. 

For this reason, strategies to reclaim time information, or mitigate against the lack of it, become important, especially because these strategies typically have limited directionality. For instance, in the crystal defect detector~\cite{Marshall:2020azl}, a prompt scintillation or phonon signal can serve as a prompt to trigger the removal of subregions in the detector bulk where the event took place. In the proposed DNA detector, this issue was not addressed~\cite{Drukier:2012hj}, although one could envision a system of microfluidics acting as a ``conveyor belt'' to transport the broken strands out of the detector. 

Obtaining timing information is the most problematic in nuclear emulsions, where the post-exposure development of the tracks is complex and time-consuming. NEWSdm propose to mount their detector and shielding on a rotating stand, so that the detector always points towards Cygnus. This strategy does not reclaim any time information, rather it removes the need for any time information by keeping the signal fixed from the perspective of the detector. In other words, if the detector is tracking Cygnus, the detector will always directly measure the recoil angle with respect to Cygnus, providing access to the DM dipole. In principle there should then be no penalty in exposure. It appears, however, that this strategy could only optimize sensitivity for one target at a time, which would seem to complicate concurrent DM searches and solar neutrino measurements. This option would also somewhat increase project cost and complexity.

We note in closing, that in the case of applied physics and calibration measurements, substantially better timing performance may be beneficial. One important example is quenching factor measurements with neutron beams, where delayed coincidence timing is used to identify matching recoil events in two detectors~\cite{Lenardo:2019vkn}. This highlights the need for smaller-scale prototypes with substantially better performance than may be required in a large, cost-optimized DM or neutrino experiment such as \Cygnus~\cite{Vahsen:2020pzb}.

\subsection{Summary of performance requirements}

In summary, we found that a directional recoil detector targeting both solar neutrinos and $\mathcal{O}(10~{\rm GeV})$ DM masses requires event-level directionality with angular resolution $\leq30^\circ$ and excellent head/tail sensitivity, ideally down to recoil energies of $\mathcal{O}(5~{\rm keV})$. A $1000~{\rm m}^3$ detector volume would require that internal electron backgrounds be reduced by factors of at least $\mathcal{O}(10^5)$, also down to $\mathcal{O}(5~{\rm keV})$. Fractional energy resolution of order 10\% at 5.9~keV appears sufficient, and even poor timing resolution, of order 0.5~h, should suffice. These requirements are consistent what was considered an optimistic performance scenario at the conclusion of a previous optimization study~\cite{Billard:2011zj}, which focused on fluorine recoils in CF$_4$. The main difference in our requirements is the need for good energy resolution, which would be needed to reconstruct neutrino spectra, but which is likely also required to achieve sufficient electron background rejection suitable for large detectors.

\section{THE CASE FOR HIGH DEFINITION RECOIL IMAGING}\label{hd_recoil_imaging}


Of all technologies on the table, gas TPCs are the closest to meeting the performance requirements we arrived at in Section~\ref{sec:performance}. Yet the optimal operating configuration in terms of gas mixture, pressure, readout segmentation, and drift length needs further study. One promising approach is high definition (HD) charge readout, meaning electronic readout with high spatial segmentation via MPGDs. High segmentation will almost certainly be required to achieve sufficient discrimination between nuclear and electron recoils. 

In the optimal case, a HD TPC would count every single electron in 3d with near unity efficiency, 100~$\upmu$m-scale segmentation, and the smallest possible diffusion---implying NID. Pixel ASIC readouts are already close to achieving this~\cite{Ligtenberg:2020ofy}, but probably not cost-effective for detectors beyond the m$^3$ scale. For larger detectors, strip readout appears more realistic, but if NID is used, this may first require development of optimized readout electronics.

As outlined in Section~\ref{sec:motivation}, there are wide applications of TPC detectors even at very small scales. A HD TPC capable of extracting all available primary ionization information would additionally enable the most precise measurements of low energy recoils possible. This will include measurements of recoil range, longitudinal and transverse straggling, and ionization quenching in gases of interest. Such measurements would allow not only validation and precise tuning of simulation tools and theoretical models for low energy nuclear recoils, but also searches for deviations---expected or otherwise---from the nominal nuclear recoil and electron recoil signatures. Such measurements would demonstrate the feasibility of directional technologies, while benefiting the whole wider field of DM detection, i.e.~including collaborations with non-directional detectors, which cannot resolve ionization distributions. Planning for HD TPC recoil measurements are already in preparation. To provide a tangible example that highlights the expected impact of this work, we end by describing one such measurement in more detail.

\subsection{The Migdal effect}\label{sec:migdal}

\begin{figure}[t]
	\includegraphics[width=\textwidth]{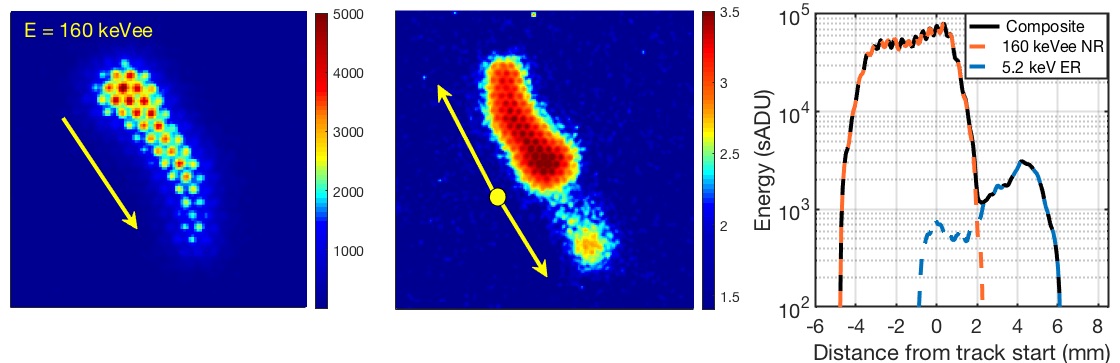}
	
	\caption{{\it Left panel}: a 160~\keVee nuclear recoil track, showing the reconstructed direction (arrow) derived from its $\mathrm{d}E/\mathrm{d}x$ profile. {\it Middle panel}: An example Migdal event constructed by taking a composite of the 160 \keVee nuclear recoil image with that of a $\sim$5.2~keV electron track with their interaction points overlaid. Due to the large difference in the dE/dx of the electron and nuclear recoil, the intensity along the electron track was scaled up by a factor 5 for visualization purposes, before co-adding to produce the image. The reconstructed directions (arrows) derived from the $\mathrm{d}E/\mathrm{d}x$ profiles are used to identify each particle in the Migdal event and its interaction point (yellow dot).
	{\it Right panel}: The $\mathrm{d}E/\mathrm{d}x$ profile of the full Migdal event. Here we show projected intensity along the major axis of the reconstructed track for both electron (blue dashed) and nuclear (orange dashed) recoils, as well as their sum (black solid). Here the true scaling between the electron and nuclear recoils was used.} 
\label{fig:MigdalAR}
\end{figure}

When performing a naive two-body nuclear scattering calculation, it is typically assumed that the electron cloud follows the recoiling nucleus instantaneously. This approximation implies that for low enough energies, at some point the resulting ionization signal is unobservably small. However, the nucleus and the atomic electron cloud are distinct entities, and taking the so-called ``Migdal approach'' of treating them as such reveals a potentially interesting new source of ionization for very low energy nuclear recoils~\cite{Baur_1983, Vegh_1983, Ruijgrok, Vergados, Sharma, Ibe:2017yqa}, as well as other detectable signals~\cite{Kouvaris:2016afs, McCabe:2017rln}. If we model the nucleus and electron cloud separately, the electrons will lag behind the nucleus during a scattering event. In the frame of the nucleus, the electron cloud is seen to experience a small boost, which can excite or ionize an electron. The effect is small but can become the dominant source of ionization at very low recoil energies. 
For example, in xenon or germanium, the maximum kinetic energy of a recoiling atom from, say, a 1 GeV DM particle would be $\sim$0.1 keV---far below experimental thresholds~\cite{Ibe:2017yqa}. Nuclear quenching will reduce the measurable energy further, compounding the problem.
Yet the Migdal prediction of the rare emission of a $\sim$keV electron would clearly be detected. So in the context of DM searches, simply invoking this effect can improve bounds for sub-GeV DM masses~\cite{Dolan:2017xbu}. Most remarkable amongst these are EDELWEISS~\cite{Armengaud:2019kfj} and XENON~\cite{Aprile:2019jmx}, who lowered their mass reach down to 45 and 85 MeV, respectively.

While calculations of the Migdal effect exist~\cite{Ibe:2017yqa, Liu:2020pat, Liang:2020ryg}, the process itself has never been measured.\footnote{The Migdal effect covers a broad range of phenomena, from $\alpha$- and $\beta$-decay, to neutron scattering. Although experiments have measured it in the former two processes \cite{Migdal-alpha,Migdal-beta1,Migdal-beta2}, they have not done so for the latter, which best approximates the light DM interaction.} This raises doubts about the validity of the effect, especially since theoretical atomic physics calculations are performed under specific assumptions, which may break down in liquids or molecular targets. A possible route towards a first experimental verification could involve a directional measurement, as has been recently proposed by the MIGDAL collaboration~\cite{MIGDALcollab}. Such a measurement would be advantageous for a conclusive identification of the effect because of the additional handle on the kinematic relationship between the Migdal electron and the recoiling nucleus that directional information provides.
Of the available directional technologies we have discussed, recoil imaging with HD gas TPCs stands out as the ideal strategy for the study of the Migdal effect. A low pressure TPC with a highly segmented ionization detector could provide both the high signal-to-noise and fine-granularity 3d track reconstruction needed to give detailed information on the low energy tracks. In contrast to DM or neutrino searches, an experiment sensitive to this rare effect (with a probability of $10^{-5} - 10^{-4}$ per nuclear recoil), would not require large volumes. Instead, one could focus on designing the best technology without the worry of scaling-up and the associated cost and complexity.



The challenge for such a measurement is to fully detect the low $\mathrm{d}E/\mathrm{d}x$ electron tracks, which requires high resolution and signal-to-noise approaching single primary electron detection. The detection of electron tracks down to a few keV has been demonstrated in Reference~\cite{Phan:2017sep} using a small TPC operating in 25--100 Torr of CF$_4$. An electron and nuclear recoil track imaged with this TPC are shown in {\bf Figure~\ref{fig:MigdalAR}}. One can see how the order of magnitude lower $\mathrm{d}E/\mathrm{d}x$ of the electron compared to the nuclear recoil could be used to distinguish them. The direction of each particle can also be deduced from the $\mathrm{d}E/\mathrm{d}x$ profile, with the nuclear recoil's falling towards the head of the track, and the electron's rising. This is a fortuitous difference that can be used to find the common vertex between them.

For a recoil imaging Migdal experiment, either optical or electronic MPGD-based readouts could work since most atoms of interest for DM searches can be found in scintillating gases. What is more important is that the detector has the highest 3d track resolution possible to measure the effect down at the energies relevant for DM searches.  In this regard, an ideal detector would be a TPC with fine-granularity MPGD readouts operating with NID.




\section{SUMMARY AND OUTLOOK}\label{sec:summary}\label{sec:outlook}

There is an emerging worldwide community interested in the directional detection of nuclear, and more recently, electron recoils. We have shown that the physics case for DM and solar neutrino recoil directionality is robust and compelling. A detector optimized for both signals would need a nuclear recoil angular resolution of order 30$^\circ$ or better, and excellent head/tail sensitivity down to recoil energies of about 5~keV. Excellent particle identification capabilities are required, which will likely require HD recoil imaging. Typical gas detector energy resolution and very modest event time resolution appear sufficient. The detector requirements to measure electron recoil signatures from neutrinos still need further study. HD recoil-imaging gas TPCs, operating at the performance limit of unity single electron efficiency and minimal achievable diffusion should also allow novel measurements of keV recoil physics, potentially including the first experimental verification of the Migdal effect. The findings would be critical to reduce simulation uncertainties and to reliably optimize the design of large-scale DM and neutrino detectors. 

\begin{summary}[SUMMARY POINTS]
\begin{enumerate}
\item A galactic dipole in direction is a robust and surprisingly model-independent signature of DM.
\item This provides powerful motivation for directional experiments, which can either directly measure the dipole, or indirectly measure it via a sidereal daily modulation. 
\item Gas TPCs and nuclear emulsions are the most advanced directional recoil detection technologies at the time of writing.
\item A ton-scale directional gas TPC could measure solar neutrinos in both nuclear and electron recoil channels while searching for DM.
\item HD nuclear recoil imaging has wide applications beyond astroparticle physics.
\item The ideal recoil-imaging gas TPC---which operates at the expected performance limits of the technology---has not yet been constructed.\vspace{-1em}
\end{enumerate}
\end{summary}
\begin{issues}[FUTURE ISSUES]
\begin{enumerate}
\item The fundamental performance limits of gas TPCs should be experimentally demonstrated: single-electron counting detector with negative ion drift (NID)
\item A single-electron counting NID TPC should be used to validate simulations of keV-scale nuclear and electron recoils
\item Simulation tools that generate the 3d topology of low-energy nuclear recoils should be developed and made publicly available.
\item Proponents of new directional technologies should demonstrate directional performance versus recoil energy.
\item The potential of liquid noble gas detectors for directional detection should be demonstrated, and compared with that of demonstrated technologies.
\item The neutrino physics potential of directional recoil detectors should be studied further and optimized in conjunction with their DM discovery capabilities.
\item The physics reach of directional electron recoil detectors should be studied further.
\item Strawman designs for directional electron recoil detectors should be developed.\vspace{-1em}
\end{enumerate}
\end{issues}

\section*{DISCLOSURE STATEMENT}
The authors are not aware of any affiliations, memberships, funding, or financial holdings that
might be perceived as affecting the objectivity of this review. 

\section*{ACKNOWLEDGMENTS}
SEV acknowledges support from the U.S. Department of Energy (DOE) via Award Number DE-SC0010504. The work of CAJO is supported by The University of Sydney. DL acknowledges support from the U.S. Department of Energy via Award Number DE-SC0019132.

\bibliographystyle{bibi}
\bibliography{biblio}

\end{document}